\documentclass[aip,graphicx]{revtex4}

\usepackage{graphicx}  
\usepackage{subfigure}
\usepackage{multirow}

\usepackage{fancyhdr}
\usepackage{longtable}
\usepackage{parskip}
\usepackage[T1]{fontenc}
\usepackage{dcolumn}   

\usepackage{bm}        
\usepackage{amsfonts}  
\usepackage{amsmath}   
\usepackage{amssymb}   

\newcommand{\pwisein}{\left\{ \begin{array}{ll}}
\newcommand{\pwiseout}{\end{array}\right.}

\let\ref\cref
\usepackage{hyperref} 
\usepackage{cleveref}

\begin{document}
\title{Anisotropic active Brownian particle in two dimensions under stochastic resetting}

\author{Anirban Ghosh}
\email{anirbansonapur@gmail.com}
\affiliation {Department of Physics, Amity Institute of Applied Sciences, Amity University, Kolkata, 700135, India}

\author{Sudipta Mandal}
\affiliation{Tata Institute of Fundamental Research, 36/P, Gopanpally Village,
Serilingampally Mandal, Ranga Reddy District, Hyderabad 500046, Telangana, India}
\affiliation{Institute of Electronic Structure and Laser, FORTH-Hellas, Greece}

\author{Subhasish Chaki}
\affiliation{Cluster of Excellence, Physics of Life, Technische Universität Dresden}
\affiliation{Institut für Theoretische Physik II—Weiche Materie, Heinrich-Heine-Universität Düsseldorf, D-40225 Düsseldorf, Germany}

\date{\today}

\begin{abstract}  

We study the dynamical behavior of an anisotropic active Brownian particle subjected to various stochastic resetting protocols in two dimensions. The motion of shape-asymmetric active Brownian particles in two dimensions leads to anisotropic diffusion at short times, whereas rotational diffusion causes the transport to become isotropic at longer times. We have considered three different resetting protocols: (a) complete resetting, when both position and orientation are reset to their initial states, (b) only position is reset to its initial state, (c) only orientation is reset to its initial state. We reveal that orientational resetting sustains anisotropy even at late times. When both the spatial position and orientation are subject to resetting, a complex position probability distribution forms in the steady state. This distribution is shaped by factors such as the initial orientation angle, the anisotropy of the particle, and the resetting rate. We have calculated the exact expressions for mean-square displacements using a renewal approach for different resetting protocols and numerically verified the analytical results. When only the translational degrees of freedom are reset, while the particle’s orientation evolves naturally, the steady-state no longer depends on particle asymmetry. In contrast, if only the orientation is reset, the long-term probability distribution becomes a Gaussian, using an effective diffusion tensor—containing non-diagonal elements—defined by the resetting rate. More broadly, the interaction between translational and rotational dynamics, in combination with stochastic resetting, produces distinct behaviors at late times that are absent in symmetric particles. Given recent progress in experimental resetting techniques, these results could be highly useful for controlling asymmetric active colloids, such as in self-assembly applications.
\end{abstract}

\maketitle 

\section{Introduction}


Active processes represent a class of dynamical systems that are inherently out of equilibrium due to their self-propelling nature\cite{ramaswamy2010mechanics,marchetti2013hydrodynamics,ten2011brownian,sandoval2013anisotropic}. Since Vicsek's pioneering work on self-propulsion\cite{vicsek1995novel}, research into active matter systems has expanded significantly, covering collective behaviors such as flocking\cite{toner2005hydrodynamics,kumar2014flocking}, clustering\cite{fily2012athermal,palacci2013living,slowman2016jamming}, and motility-induced phase separation (MIPS)\cite{Gompper_2020,schwarz2012phase,redner2013structure,stenhammar2015activity}. Theoretical approaches to understanding active matter often center around the analysis of simplified yet analytically solvable models, such as the Run-and-Tumble Particle (RTP) and the Active Brownian Particle (ABP), as well as various extensions and modifications of these foundational models\cite{siegle2010origin,ten2011brownian,santra2022universal,li2017two,ao2014active,solon2015active,shee2020active,basu2019long,tailleur2008statistical,santra2021directional,ghosh2025finite}. In these models, activity emerges from the interaction between spatial motion and a stochastic internal orientational degree of freedom. The intrinsic timescale linked to the particle's internal orientation gives rise to a variety of fascinating phenomena, even at the level of an individual particle, such as spatial anisotropy and ballistic motion over short durations\cite{mandal2024diffusion,ten2011brownian,PhysRevE.98.062121}.

Numerous processes in physics, chemistry, and biology involve diffusing particles with highly anisotropic shapes. Notable instances contain rod-shaped bacteria or viruses\cite{bawden1936liquid,wen1989observation,graf1999phase,doostmohammadi2016defect}, anisotropic colloids\cite{lowen1994brownian,bolhuis1997tracing,dijkstra2014entropy,lowen1999anisotropic,chaki2025dynamics}, and nematic macromolecules\cite{gray1984theory,narayan2007long}. Such shape anisotropy occurs across a wide range of scales, from a few nanometers to several micrometers. Analyzing the diffusive behavior of anisotropic particles is naturally more complicated than that of spherical particles, primarily due to the coupling between rotational and translational motions. In the short-time regime, the anisotropic behavior emerges from the correlation between the particle's instantaneous orientation and its translational motion\cite{grima2007brownian,ghosh2020persistence}. The motion of anisotropic Active Brownian Particles (ABPs) is marked by a transition from the initial anisotropic diffusion, driven by the particle's starting orientation, to an effective isotropic diffusion at longer times.

Stochastic resetting involves intermittently interrupting and restarting a dynamical process. The inclusion of a resetting mechanism in a stochastic system can dramatically change both its static and dynamic properties\cite{evans2020stochastic}. This concept is relevant across diverse fields, such as search problems\cite{benichou2011intermittent,chechkin2018random,pal2017first,pal2020search,sar2023resetting,paramanick2024uncovering}, population dynamics\cite{manrubia1999stochastic,visco2010switching}, computer science\cite{montanari2002optimizing}, and biological systems\cite{kussell2005phenotypic,kussell2005bacterial,roldan2016stochastic}. A well-known example of stochastic resetting is a Brownian particle undergoing diffusion that is periodically reset to its initial location at a specific rate\cite{evans2011diffusion}. This resetting process drives the system away from equilibrium, resulting in intriguing outcomes like non-equilibrium steady states, unique temporal relaxation behaviors, and non-monotonic mean first passage times\cite{evans2011diffusion,evans2012diffusion,majumdar2015dynamical}. The impact of resetting on various diffusive processes has been extensively studied in the past decade\cite{santra2020run,olsen2023steady,radice2021one,stojkoski2022autocorrelation,chatterjee2018diffusion,PhysRevE.109.024134}. While it might seem intuitive that restarting a stochastic process repeatedly would delay its overall completion, this is not necessarily the case. In fact, a surprising feature of stochastic resetting is its potential to speed up the completion of specific random search processes. As an instance, resetting can lower the mean first-passage time for a diffusive particle to reach a target\cite{reuveni2016optimal,evans2011diffusion,pal2017first,tal2020experimental,besga2020optimal,faisant2021optimal}, decrease the runtime of stochastic algorithms\cite{lorenz2021restart,lorenz2018runtime}, and reduce the turnover time in enzymatic reactions\cite{reuveni2014role,rotbart2015michaelis}. Therefore, stochastic resetting is an effective method for optimizing the completion times of stochastic processes and dynamical systems.

While most previous studies on resetting have focused on symmetric particles, this paper investigates the impact of resetting on an anisotropic active Brownian particle (ABP) in two dimensions. The ABP in our study is an overdamped particle with an orientational degree of freedom, undergoing rotational diffusion. Its behavior in two-dimensional space is defined by its position $(x, y)$ and orientation $\theta$. We examine three distinct resetting protocols: (a) complete resetting, where both the particle's position and orientation are reset to their initial values at a rate $r$, (b) resetting of only the position, and (c) resetting of only the orientation angle. In the first two protocols involving position resetting, the particle's mean-square displacement (MSD) reaches a steady state. Conversely, for the protocol that resets only the orientation, the particle does not attain a stationary state. Instead, it exhibits anisotropic diffusion, with a transition from ballistic to diffusive behavior as time progresses. The structure of the paper is as follows: Sec.(\ref{model}) introduces the fundamental model and theoretical framework; Sec.(\ref{complete_reset}) elaborates on the complete resetting protocol, detailing its dynamics; Sec.(\ref{position_reset}) focuses on the dynamics under the position-reset protocol; Sec.(\ref{angle_reset}) discusses the effects of orientation resetting; Sec.(\ref{sec:conclusions}) presents the conclusions. Additionally, Appendix(\ref{perturbative}) provides a comprehensive analytical derivation of steady state probability distribution function using a perturbative scheme.

\section{Model and Theory}\label{model}
\begin{figure}
    \centering
    \includegraphics[width=1\textwidth]{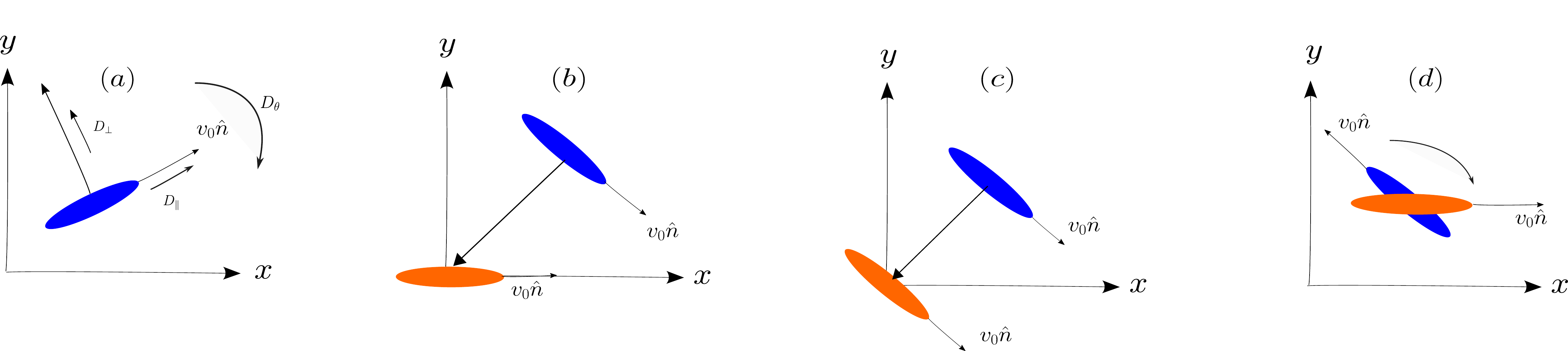}
    \caption{(a) The basic model of an anisotropic ABP incorporates parallel, perpendicular, and rotational diffusion coefficients, with the propulsion velocity $v_0$ directed along the particle's major axis. (b) In the position-orientation complete resetting protocol, both the particle's position and orientation are reset simultaneously. (c) The position-only resetting protocol involves resetting the particle's position to the origin while leaving its orientation unchanged. (d) The orientation-only resetting protocol resets the particle's orientation to its initial orientation as $\theta=0$ without altering its position.}
    \label{reset_figure}
\end{figure}

We have investigated the dynamics of an anisotropic active Brownian particle (ABP) subjected to resetting. This particle is confined to move within a two-dimensional framework and possesses a propulsion velocity along its longer axis, denoted as ($\Tilde{x}$). The particle's orientation in the $x$-$y$ plane plays a crucial role in its motion. This orientation can be characterized by a unit vector $\hat{n}_i=(\cos{\theta_i},\sin{\theta_i})$, which defines the direction of the propulsion velocity. In the lab frame ($x$-$y$ plane), the particle's state at any given time $t$ can be described by the position vector $R(t)$, which indicates the coordinates of its center of mass. This vector also represents the coordinates relative to the particle’s intrinsic body frame $(\delta\Tilde{x},\delta\Tilde{y})$. The angle $\theta(t)$ represents the orientation of the particle, specifically the angle between the $x$ axis of the laboratory frame and the $\Tilde{x}$ axis of the body frame. Understanding the frame of reference is essential for analyzing the behavior of anisotropic particles, as the rotational and translational motions in the body frame are decoupled. The motion of the anisotropic ABP as shown in fig.(\ref{reset_figure}a) is governed by the Langevin equation in body frame, which can be expressed as:

\begin{equation}
\begin{split}
&\frac{1}{\Gamma_{\parallel}}\frac{\partial \Tilde{x}}{\partial t}=\frac{v_0}{\Gamma_{\parallel}}+\Tilde{\eta}_x(t)\\
&\frac{1}{\Gamma_{\perp}}\frac{\partial \Tilde{y}}{\partial t}=\Tilde{\eta}_y(t)\\
&\frac{1}{\Gamma_{\theta}}\frac{\partial \Tilde{\theta}}{\partial t}=\Tilde{\eta}_\theta(t)
\end{split}
\label{1}
\end{equation}

Here $v_0$ is the propulsion velocity in the body frame, it is directed along the longer axis of the anisotropic particle. The mobilities along its longer and shorter axes are respectively $\Gamma_{\parallel}=\frac{D_{\parallel}}{k_BT}$ and $\Gamma_{\perp}=\frac{D_{\perp}}{k_BT}$. $\Gamma_{\theta}=\frac{D_{\theta}}{k_BT}$ is the rotational mobility. Here $D_{\parallel}$, $D_{\perp}$ and $D_{\theta}$ represent diffusion coefficients along the parallel, perpendicular and rotational axes respectively. The correlations of the thermal fluctuations in the body frame are described as,
\begin{equation}
\begin{split}
&\langle\Tilde{\eta}\rangle=0\\
&\langle\Tilde{\eta}_i(t)\Tilde{\eta}_j(t^\prime)\rangle=2k_BT\Gamma_i\delta_{ij}\delta(t-t^\prime)
\end{split}
\end{equation}

To derive these equations in the laboratory frame, we must perform a straightforward coordinate rotation. The displacement in the two frames is related with the following equations;

\begin{equation}
\begin{split}
&\delta x= \cos\theta\delta\Tilde{x}-\sin\theta\delta\Tilde{y}\\
&\delta y= \sin\theta\delta\Tilde{x}+\cos\theta\delta\Tilde{y}
\end{split}
\end{equation}

Taking the limit $\delta t\rightarrow 0$ after dividing the equations by $\delta t$ and substituting the expressions for linear and angular velocities in the body frame from Eq.(\ref{1}), we derive the final set of equations that describe the motion in the lab frame,

\begin{equation}
\begin{split}
&\frac{\partial x}{\partial t}=v_0\cos{\theta}(t)+\eta_x(t)\\
&\frac{\partial y}{\partial t}=v_0\sin{\theta}(t)+\eta_y(t)\\
&\frac{\partial \theta}{\partial t}=\eta_\theta(t)
\end{split}
\label{lab}
\end{equation}
The random noise $\eta_i$ have zero mean and the correlations at fixed $\theta(t)$ are given by

\begin{equation}
\begin{split}
&\langle\eta_i(t)\eta_j(t^\prime)\rangle=2k_BT\Gamma_{ij}[\theta(t)]\delta(t-t^\prime)\\
&\langle\eta_\theta(t)\eta_\theta(t^\prime)\rangle=2k_BT\Gamma_\theta\delta(t-t^\prime)
\end{split}
\end{equation}

Here 

\begin{equation}
\Gamma_{ij}[\theta(t)]=\bar{\Gamma}\delta_{ij}+\frac{\Delta\Gamma}{2}
\begin{pmatrix}
\cos2\theta & \sin2\theta \\ \sin2\theta & -\cos2\theta
\end{pmatrix}
\end{equation}
where $\Bar{\Gamma}=\frac{(\Gamma_{\parallel}+\Gamma_{\perp})}{2}$ and $\Delta\Gamma=(\Gamma_{\parallel}-\Gamma_{\perp})$. The average diffusion coefficient is also defined as $\Bar{D}=k_BT\Bar{\Gamma}=\frac{(D_\parallel+D_\perp)}{2}$ and asymmetry factor difference between the two diffusion coefficients along parallel and perpendicular axes is defined as $\Delta D=k_BT\Delta\Gamma=(D_\parallel-D_\perp)$. 

We integrate Eq.(\ref{lab}) with respect to $t$ for $x$, and it is found as,
\begin{equation}
x(t)=\int_{0}^{t}dt^\prime\eta_x(t^\prime)+\int_{0}^{t}dt^\prime v_0\cos{\theta}(t^\prime)
\end{equation}

The second moment of the particle along $x$ and $y$ axes with an initial angle $\theta_0$ without any resetting, is calculated as\cite{mandal2024diffusion,ghosh2022persistence}

\begin{widetext}
\begin{equation}
\langle x^2(t)\rangle_{0}=\Bigg[2\bar{D}t+\Delta D\cos 2\theta_0\Big(\frac{1-e^{-4D_\theta t}}{4D_\theta}\Big)\Bigg]+\frac{v_0^2\cos 2\theta_0}{12D_\theta^2}\big(3-4e^{-D_\theta t}+e^{-4D_\theta t}\big)+\frac{v_0^2}{D_\theta^2}\big(D_\theta t+e^{-D_\theta t}-1\big)
\label{eqn:msd1}
\end{equation}

\begin{equation}
\langle y^2(t)\rangle_{0}=\Bigg[2\bar{D}t-\Delta D\cos 2\theta_0\Big(\frac{1-e^{-4D_\theta t}}{4D_\theta}\Big)\Bigg]-\frac{v_0^2\cos 2\theta_0}{12D_\theta^2}\big(3-4e^{-D_\theta t}+e^{-4D_\theta t}\big)+\frac{v_0^2}{D_\theta^2}\big(D_\theta t+e^{-D_\theta t}-1\big)
\label{eqn:msd2}
\end{equation}
\end{widetext}
We anticipate that the interaction between the two timescales, \( D_\theta^{-1} \) and the resetting rate inverse \( r^{-1} \), will give rise to rich dynamics in the case of an anisotropic ABP under resetting, given that both \( D_\theta \) and \( r \) have units of inverse time. In the long time regime when $t\gg D_\theta^{-1}$, the mean-square displacements along both the $x$ and $y$ axes exhibit diffusive behavior similar to an isotropic particle expressed as\cite{mandal2024diffusion,ghosh2022persistence},

\begin{equation}
\begin{split}
    &\langle\Delta x^2(t)\rangle_{0}=2\Bigg(\Bar{D}+\frac{v_0^2}{2D_\theta}\Bigg)t+\frac{\Delta D}{4D_\theta}-\frac{7v_0^2}{4D_\theta^2}\\
    &\langle\Delta y^2(t)\rangle_{0}=2\Bigg(\Bar{D}+\frac{v_0^2}{2D_\theta}\Bigg)t-\frac{\Delta D}{4D_\theta}-\frac{5v_0^2}{4D_\theta^2}
\end{split}
\label{msdlong}
\end{equation}

Furthermore, at the short time regime $t< D_\theta^{-1}$, we find initially instantaneous diffusion regime and then super-diffusion regime as depicted\cite{mandal2024diffusion}
\begin{equation}
\begin{split}
&\langle\Delta x^2(t)\rangle_0=2D_{\parallel}t-2D_\theta\Delta Dt^2+\mathcal{O}(t^3)\\
&\langle\Delta y^2(t)\rangle_0=2D_{\perp}t+2D_\theta\Delta D t^2+\mathcal{O}(t^3)
\end{split}
\label{small_D}
\end{equation}

In this paper, we study the effect of stochastic resetting on the dynamics of and anisotropic ABP. For ABP, the presence of rotational and translational degrees of freedom give rise to the chance of resetting in the phase space, instead of in position space only. The first explored study shows different resetting protocols involving position and velocity in one-dimensional RTP\cite{evans2018run}. In the case of anisotropic ABP, where the translational and rotational degrees are the important variables, the similar situation arises like RTP. In the following, we study three different resetting protocols.

(I) \textbf{Position-orientation resetting}: All the phase variables are reset to its initial values with the reset rate $r$ as shown in the fig.(\ref{reset_figure}b). We assumed that the particle starts from the origin and initial angle $\theta_0=0$. The particle resets to $x=y=0$ and $\theta_0=0$ at any time $t$ with reset rate $r$.

(II) \textbf{Position resetting}: According to this protocol shown in the fig.(\ref{reset_figure}c), we reset the position of the particle to the origin with rate $r$, but the orientation is not reset. The orientation evolves according to the normal diffusion dynamics.

(III) \textbf{Orientation resetting}: In this protocol shown in the fig.(\ref{reset_figure}d), only the orientation $\theta$ resets to $\theta=0$ with rate $r$, and the position is not reset. 

\section{Anisotropic ABP with resetting} \label{reset}
\subsection{Position-orientation Resetting}\label{complete_reset}
    The most common resetting protocol occurs when both the position and the orientation of the particle are reset to their initial values at a rate \( r \).
 Our focus lies on the position distribution \( P(x,y,t) = \int d\theta \, \mathcal{P}(x,y,\theta,t) \), which is obtained by integrating \( \mathcal{P}(x,y,\theta,t) \), the probability density of the particle being at position \( (x,y) \) with orientation \( \theta \) at time \( t \), over all possible orientations \( \theta \).
A renewal equation for \(\mathcal{P}(x,y,\theta,t)\), the joint probability density of the particle's position and orientation, can be written directly as:

    \begin{equation}
        \mathcal{P}(x,y,\theta,t)=e^{-rt}\mathcal{P}_0(x,y,\theta,t)+r\int_{0}^{t}d\tau e^{-r\tau}\mathcal{P}_0(x,y,\theta,\tau)
    \end{equation}
   Here, \(\mathcal{P}_0(x,y,\theta,t)\) represents the probability that, in the absence of resetting, the active Brownian particle (ABP) is located at position \((x,y)\) with orientation \(\theta\) at time \(t\), starting from the initial position \(x=0, y=0\). The renewal equation consists of two terms: 

The first term accounts for the scenario where no resetting events occur up to time \(t\). The second term captures the contribution from trajectories where the most recent resetting event occurred at a time \(t-\tau\) prior.

    A corresponding renewal equation governing the position distribution can be formulated by integrating over the orientation \(\theta\), as follows:

    \begin{equation}
        P(x,y,t)=e^{-rt}P_0(x,y,t)+r\int_{0}^{t}d\tau e^{-r\tau}P_0(x,y,\tau)
        \label{12}
    \end{equation}

    The model investigated in the work of Kumar et. al\cite{kumar2020active} is different than the one that is presented here. They considered an active Brownian particle without the thermal fluctuations and an explicit shape anisotropy. The model is described in a frame fixed to the particle, the translational and the rotational motion of the particle is completely decoupled. However, in our model in lab-frame, the shape asymmetry of the particle leads to a coupling between the translational and rotational motions of the particle.

    \subsubsection{Moments}
    To understand the influence of position-orientation resetting on the dynamical behavior of the anisotropic ABP, we examine the moments of the position coordinates. Clearly, under the influence of resetting, the moments adhere to a renewal equation analogous to Eq.(\ref{12}). As an illustration, we can multiply both sides by $x^n$ and integrate with respect to both $x$ and $y$, which leads to the resulting expression:

 \begin{equation}
 \begin{split}
      &  \langle x^n(t)\rangle=e^{-rt}\langle x^n(t)\rangle_{0}+r\int_{0}^{t}d\tau e^{-r\tau}\langle x^n(\tau)\rangle_{0}\\
      &\langle y^n(t)\rangle=e^{-rt}\langle y^n(t)\rangle_{0}+r\int_{0}^{t}d\tau e^{-r\tau}\langle y^n(\tau)\rangle_{0}
        \end{split}
        \label{ren}
    \end{equation}
   Here, $\langle x^n(t)\rangle_{0}$ is the $n$th moment of the $x$-coordinate of the position in the absence of any resetting events, and it can be computed explicitly for any value of $n$. 

In the following, we will derive the first two moments for the $x$ and $y$ components, utilizing the established expressions for a free ellipsoidal Active Brownian Particle (ABP). The average values of the $x$ and $y$ coordinates are given by:

\begin{equation}
    \begin{split}
        &\langle x(t)\rangle_0^{\theta_0}=v_0\int_{0}^{t}d\tau \langle\cos{\theta(\tau)}\rangle_0^{\theta_0}+\int_{0}^{t}\langle\eta_x(\tau)\rangle d\tau\\
        &\langle y(t)\rangle_0^{\theta_0}=v_0\int_{0}^{t}d\tau \langle\sin{\theta(\tau)}\rangle_0^{\theta_0}+\int_{0}^{t}\langle\eta_y(\tau)\rangle d\tau
    \end{split}
    \label{eq1}
\end{equation}
where we have used the superscript $\theta_0$ to denote the initial orientation.
\begin{equation}
    \begin{split}
        &\langle\cos{\theta(\tau)}\rangle_0^{\theta_0}=\cos{\theta_0}e^{-D_\theta \tau}\\
        &\langle\sin{\theta(\tau)}\rangle_0^{\theta_0}=\sin{\theta_0}e^{-D_\theta \tau}
    \end{split}
\end{equation}
The average positions can be computed using the above equations in Eq(\ref{eq1}),
\begin{equation}
    \begin{split}
        &\langle x(t)\rangle_0^{\theta_0}=\frac{v_0\cos{\theta_0}}{D_\theta}(1-e^{-D_\theta t})\\
        &\langle y(t)\rangle_0^{\theta_0}=\frac{v_0\sin{\theta_0}}{D_\theta}(1-e^{-D_\theta t})
    \end{split}
    \label{eq2}
\end{equation}
Using Eq.(\ref{ren}) we get the first moment after resetting at $\theta_0=0$
\begin{equation}
    \langle x(t)\rangle=\frac{v_0}{r+D_\theta}(1-e^{-(r+D_\theta)t})
    \label{eq3}
\end{equation}
while $\langle y(t)\rangle=0$ for every value of time. This observation indicates the emergence of a timescale due to the resetting mechanism. At short times, specifically for $t \ll (r + D_\theta)^{-1}$, the particle exhibits motion along the $x$-axis with an effective velocity $v_0$, similar to the behavior of a free anisotropic Active Brownian Particle (ABP). On the other hand, at longer times, the particle approaches a stationary position, with its mean position shifting closer to the origin as the resetting rate $r$ increases.

The second moments of $x$ and $y$ without resetting of ellipsoidal ABP are given in Eqs.(\ref{eqn:msd1}) and (\ref{eqn:msd2}).
Next, we compute the second moments using Eq.(\ref{ren}), Eq.(\ref{eqn:msd1}), and Eq.(\ref{eqn:msd2}) with $n=2$. Here, we investigate the behavior of the mean squared displacement (MSD), defined as $\sigma_x^2(t) = \langle x^2(t) \rangle - \langle x(t) \rangle^2$ for the $x$-component and $\sigma_y^2(t) = \langle y^2(t) \rangle$ for the $y$-component. We obtain precise analytical formulas for the higher-order moments of the $x$ and $y$ components of the position under the position-orientation resetting protocol. The second moment of $x(t)$ and $y(t)$ can be explicitly calculated from Eq.(\ref{ren}) as

\begin{widetext}
\begin{equation}
    \begin{split}
        \langle x^2(t)\rangle&=\frac{2\Bar{D}}{r}+\frac{\Delta D\cos{2\theta_0}}{r+4D_\theta}+\frac{v_0^2}{r(r+D_\theta)}+\frac{v_0^2\cos{2\theta_0}}{(r+D_\theta)(r+4D_\theta)}+e^{-rt}\Bigg[\frac{v_0^2\cos{2\theta_0}e^{-4D_\theta t}}{3D_\theta(r+4D_\theta)}-\frac{v_0^2\cos{2\theta_0}e^{-D_\theta t}}{3D_\theta(r+D_\theta)}-\frac{v_0^2}{rD_\theta}\\
        &-\frac{2\Bar{D}}{r}+\frac{v_0^2e^{-D_\theta t}}{D_\theta(r+D_\theta)}-\frac{\Delta D\cos{2\theta_0}e^{-4D_\theta t}}{r+4D_\theta}\Bigg]
    \end{split}
    \label{msdxreset}
\end{equation}

\begin{equation}
    \begin{split}
        \langle y^2(t)\rangle&=\frac{2\Bar{D}}{r}-\frac{\Delta D\cos{2\theta_0}}{r+4D_\theta}+\frac{v_0^2}{r(r+D_\theta)}-\frac{v_0^2\cos{2\theta_0}}{(r+D_\theta)(r+4D_\theta)}+e^{-rt}\Bigg[\frac{\Delta D\cos{2\theta_0}e^{-4D_\theta t}}{r+4D_\theta}+\frac{v_0^2\cos{2\theta_0}e^{-D_\theta t}}{3D_\theta(r+D_\theta)}\\
        &-\frac{v_0^2}{rD_\theta}-\frac{2\Bar{D}}{r}-\frac{v_0^2\cos{2\theta_0}e^{-4D_\theta t}}{3D_\theta(r+4D_\theta)}+\frac{v_0^2e^{-D_\theta t}}{D_\theta(r+D_\theta)}\Bigg]
    \end{split}
    \label{msdyreset}
\end{equation}
\end{widetext}
For $\theta_0 = 0$, where the particle is initially oriented along the $x$ axis, the mean-squared displacements along the $x$ and $y$ directions are given as:

\begin{widetext}
\begin{equation}
    \begin{split}
        \sigma_x^2(t)&=\frac{2\Bar{D}}{r}+\frac{\Delta D}{r+4D_\theta}+\frac{2v_0^2(r+2D_\theta)}{r(r+D_\theta)(r+4D_\theta)}+e^{-rt}\Bigg[\frac{2v_0^2e^{-D_\theta t}}{3D_\theta(r+D_\theta)}+\frac{e^{-4D_\theta t}}{r+4D_\theta}\Big(\frac{v_0^2}{3D_\theta}-\Delta D\Big)-\frac{v_0^2}{rD_\theta}-\frac{2\Bar{D}}{r}\Bigg]\\
        &-\frac{v_0^2}{(r+D_\theta)^2}\Big(1-e^{-(r+D_\theta)t}\Big)^2
    \end{split}
    \label{msdx}
\end{equation}

\begin{equation}
    \begin{split}
        \sigma_y^2(t)&=\frac{2\Bar{D}}{r}-\frac{\Delta D}{r+4D_\theta}+\frac{4D_\theta v_0^2}{r(r+D_\theta)(r+4D_\theta)}+e^{-rt}\Bigg[\frac{4v_0^2e^{-D_\theta t}}{3D_\theta(r+D_\theta)}+\frac{e^{-4D_\theta t}}{r+4D_\theta}\Big(\Delta D-\frac{v_0^2}{3D_\theta}\Big)-\frac{v_0^2}{rD_\theta}-\frac{2\Bar{D}}{r}\Bigg]
    \end{split}
    \label{msdy}
\end{equation}
\end{widetext}

\begin{figure}
    \centering
    \includegraphics[width=1\textwidth]{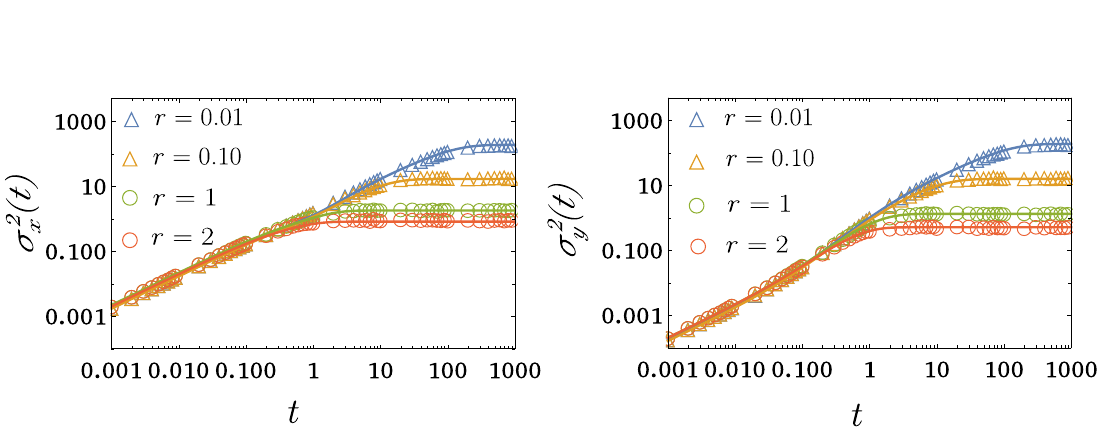}
    \caption{Position-orientation resetting: Mean squared displacements along $x$ and $y$ directions as a function of time $t$ for $D_\theta=1$, $D_\parallel=1$, $D_\perp=0.1$, $v_0=1$, $\theta_0=0$ and varying resetting rates $r$. The simulation data are represented by symbols, while the solid curves correspond to the analytical predictions from Eqs.(\ref{msdx}) and (\ref{msdy}). The stationary values decrease as the resetting rate $r$ increases.}
    \label{fig2}
\end{figure}

In the transient time regime, we can expand the exponential terms in Eqs.~(\ref{msdx}) and (\ref{msdy}) up to the first order in \( t \). In this regime, both \( \sigma_x^2 \) and \( \sigma_y^2 \) are independent of the resetting rate and exhibit purely diffusive behavior. Their asymptotic forms for \( t \to 0 \) are given by:

\begin{align*}
    \sigma_x^2(t \to 0) &= 2D_\parallel t, \\
    \sigma_y^2(t \to 0) &= 2D_\perp t.
\end{align*}

This result highlights that, at very short times, the system follows normal diffusion before any significant effects of resetting come into play. This result differs significantly from the work of Kumar et al.\cite{kumar2020active}, as their model does not include the thermal fluctuation term in the absence of anisotropy effects. In contrast, our model explicitly incorporates the fluctuation term along with anisotropy. This thermal fluctuation term plays a crucial role in governing the transient diffusive dynamics, which are strongly influenced by anisotropy.

If we expand the exponential terms up to second order, we find

\begin{widetext}
\begin{equation}
\begin{split}
  \sigma_x^2(t)&=\Big(2D_{\parallel}-\frac{v_0^2}{r+D_\theta}\Big)t-\Bigg(rD_{\parallel}+\frac{v_0^2(2r-11D_\theta)}{6D_\theta}+2D_\theta\Delta D\Bigg)t^2 +\Bigg(2rD_\theta\Delta D-\frac{rv_0^2(3r+4D_\theta)}{3(r+D_\theta)}\Bigg)t^3\\
    &+\Bigg(\frac{r^2v_0^2D_\theta(5r+8D_\theta)}{(r+D_\theta)(r+4D_\theta)}-\frac{4r^2D_\theta^2\Delta D}{r+4D_\theta}\Bigg)t^4+\mathcal{O}(t^5)
   \end{split}
   \label{complete_shortx}
\end{equation}

\begin{equation}
\begin{split}
    \sigma_y^2(t)=2D_\perp t+\Big(2D_\theta\Delta D-rD_{\perp}\Big)t^2-2rD_\theta\Delta Dt^3+\Bigg(\frac{4r^2D_\theta^2\Delta D}{r+4D_\theta}-\frac{r^3v_0^2}{(r+D_\theta)(r+4D_\theta)}\Bigg)t^4+\mathcal{O}(t^5)
   \end{split}
   \label{complete_shorty}
\end{equation}
\end{widetext}

At short times, Eqs.~(\ref{complete_shortx}) and (\ref{complete_shorty}) indicate that the system exhibits super-diffusive behavior, emphasizing the pronounced anisotropy in the dynamics. This occurs after a very brief transient phase of diffusive dynamics, which remains independent of the reset rate \( r \). 

It is intriguing to contrast this short-time behavior with that of an ellipsoidal active Brownian particle (ABP) without resetting. When starting from the origin and aligned along the $x$ axis, the mean squared displacement (MSD) of a regular ellipsoidal ABP along the $x$ axis exhibits hyper-diffusive behavior in the short-time regime\cite{mandal2024diffusion}. In the presence of position-orientation resetting, however, both $\sigma_x^2$ and $\sigma_y^2$ exhibit growth with a super-diffusive nature. Additionally, the resetting significantly alters the leading-order behavior of the mean squared displacement (MSD) as shown in the fig.(\ref{fig2}).

At long times, the particle is anticipated to reach a stationary state in which the mean squared displacement (MSD) becomes time-independent and attains a constant value.

\begin{equation}
    \begin{split}
    &\lim_{t\rightarrow\infty}\sigma_x^2=\frac{2\Bar{D}}{r}+\frac{\Delta D}{r+4D_\theta}+\frac{v_0^2(r^2+2rD_\theta+4D_\theta^2)}{r(r+D_\theta)^2(r+4D_\theta)}\\
    &\lim_{t\rightarrow\infty}\sigma_y^2=\frac{2\Bar{D}}{r}-\frac{\Delta D}{r+4D_\theta}+\frac{4D_\theta v_0^2}{r(r+D_\theta)(r+4D_\theta)}
    \end{split}
\end{equation}

\begin{figure}
    \centering
    \includegraphics[width=1\linewidth]{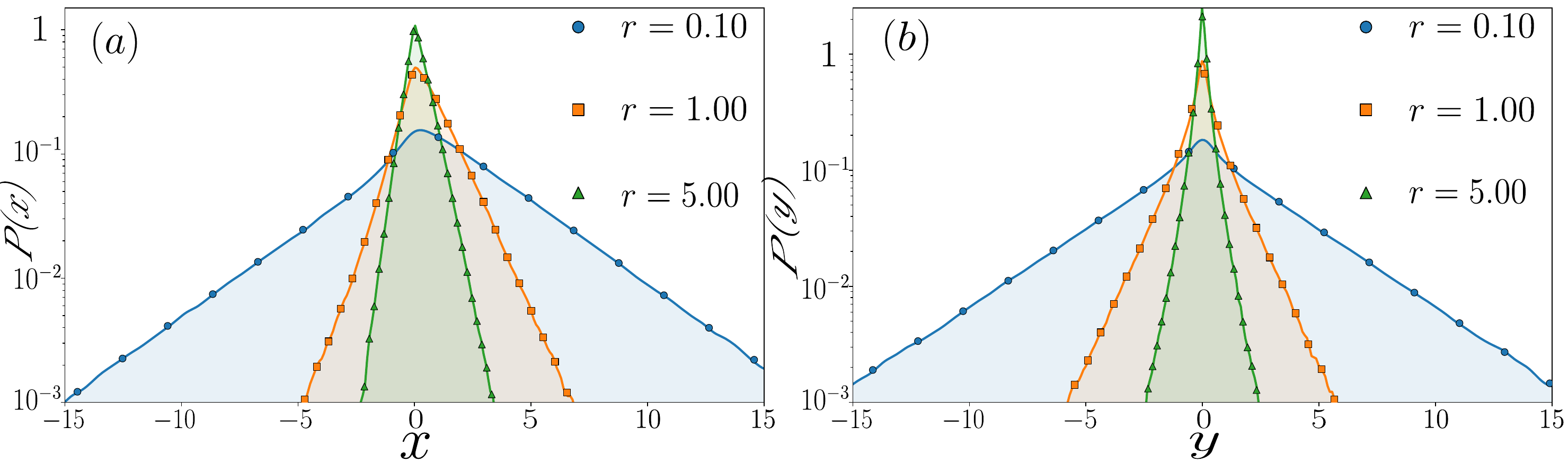}
    \caption{Steady state position distribution along $x$ and $y$ directions of anisotropic ABP under position-orientation resetting when $D_\theta=0.1$, $D_\parallel=1$, $D_\perp=0.1$, $v_0=1$ and $\theta_0=0$ for different values of reset rate $r$ as shown in the legends taken at $t=100$.}
    \label{fig3}
\end{figure}

Interestingly, the stationary mean squared displacement (MSD) exhibits distinct values for the $x$ and $y$ components, reflecting the continued presence of anisotropy. This outcome is anticipated, as resetting to $\theta_0 = 0$ consistently introduces a pronounced directional bias with each resetting event. Fig.(\ref{fig2}) illustrate the plots of $\sigma_x^2$ and $\sigma_y^2$ as functions of time $t$ for various values of $r$. As expected, the MSD saturates faster to its stationary value with increasing $r$. The leading-order time-scale of relaxation to the stationary value is $\tau_{relax}^{pos+ori}=\min[(r+D_\theta)^{-1},(r+4D_\theta)^{-1}]$.

Figs.(\ref{fig3}(a)) and (\ref{fig3}(b)) show the simulation results for the long time position distribution for both $x$ and $y$. For a small resetting rate, the particle behaves like a free particle for an extended duration before being reset to the origin. Conversely, for a large resetting rate, the time interval between successive resetting events is significantly shorter than the rotational diffusion timescale of the free anisotropic ABP dynamics. The anisotropic characteristics of the position distribution, as depicted in figs.(\ref{fig3}(a)) and (\ref{fig3}(b)), reveal that the distribution along the $y$-direction significantly differs from that along the $x$-direction.

A novel perturbative scheme is employed to analytically determine the distribution function of the anisotropic active Brownian particle (ABP). The complete calculation and details of the method are provided in Appendix \ref{perturbative}.

\subsection{Position Resetting}\label{position_reset}

In this section, we analyze the dynamics of the active Brownian particle (ABP) under resetting protocol II, wherein only the particle's position is reset. Specifically, the particle is returned to the origin, $x = y = 0$, at a resetting rate $r$, while its orientation $\theta$ remains unaffected by the reset. As in the earlier case, the particle begins its motion at the origin with an initial orientation $\theta = 0$ at time $t = 0$. Consequently, the distribution of $\theta$ at any time $t$ remains Gaussian in nature, with a zero mean and a variance of $2D_\theta t$. The corresponding renewal equation is formulated as follows:

\begin{equation}
\begin{split}
P_r(x,y,\theta,t|0,0,\theta_0)&=e^{-rt}P_0(x,y,\theta,t|0,0,\theta_0)+r\int_{0}^{t}d\tau e^{-r\tau}\int dx^\prime \int dy^\prime \int d\theta^\prime P_r(x^\prime, y^\prime, \theta^\prime, t-\tau|0, 0, \theta_0)\times \\
&P_0(x,y,\theta, \tau|0,0, \theta^\prime)
\end{split}
\end{equation}

The propagator for $x$ in this scenario can be derived by integrating the previously stated equation over the variables $y$, $\theta$, $x^\prime$, and $y^\prime$, as follows:

\begin{equation}
\begin{split}
P_{pr}(x,t|0,\theta_0)=e^{-rt}P_0(x,t|0,\theta_0)+r\int_{0}^{t}d\tau e^{-r\tau}\int d\theta^\prime P_0(\theta^\prime,t-\tau|\theta_0)P_0(x,\tau|0,\theta^\prime)
\end{split}
\end{equation}

The corresponding $n$-th moment for $x$ in this scenario can be expressed as:

\begin{equation}
\begin{split}
\langle x^n(t)|\theta_0\rangle_{pr}=e^{-rt}\langle x^n(t)|\theta_0\rangle_0+r\int_{0}^{t}d\tau e^{-r\tau}\int d\theta^\prime P_0(\theta^\prime,t-\tau|\theta_0)\langle x^n(\tau)|\theta^\prime\rangle_0
\end{split}
\label{1000}
\end{equation}
Now, we have calculated the second moment of $x$ for the anisotropic ABP putting $n=2$ in Eq.(\ref{1000}) and extracting $\langle x^2(t)|\theta_0\rangle_0$ from Eq.(\ref{eqn:msd1})

\begin{widetext}
\begin{equation}
\begin{split}
\langle x^2(t)|\theta_0\rangle_{pr}&=e^{-rt}\Big[2\Bar{D}t+\frac{\Delta D\cos{2\theta_0}}{4D_\theta}(1-e^{-4D_\theta t})+\frac{v_0^2\cos{2\theta_0}}{12D_\theta^2}(3-4e^{-D_\theta t}+e^{-4D_\theta t})+\frac{v_0^2}{D_\theta^2}(D_\theta t+e^{-D_\theta t}-1)\Big]\\
&+r\int_{0}^{t}d\tau e^{-r\tau}\int d\theta^\prime\frac{e^{-\frac{(\theta^\prime-\theta_0)^2}{4D_\theta(t-\tau)}}}{\sqrt{4\pi D_\theta(t-\tau)}}\Big[2\Bar{D}\tau+\frac{\Delta D\cos{2\theta^\prime}}{4D_\theta}(1-e^{-4D_\theta\tau})+\frac{v_0^2\cos{2\theta_0}}{12D_\theta^2}(3-4e^{-D_\theta\tau}+e^{-4D_\theta\tau})\\
&+\frac{v_0^2}{D_\theta^2}(D_\theta\tau+e^{-D_\theta\tau}-1)\Big]
\end{split}
\label{pos_reset_renewal}
\end{equation}
\end{widetext}

We will use the following properties to calculate the above integral

\begin{equation}
\begin{split}
 \langle\cos{2\theta(t)}|\theta_0\rangle&=Re\Bigg[\int d\theta e^{i2\theta}\frac{e^{-\frac{(\theta-\theta_0)^2}{4D_\theta(t-\tau)}}}{\sqrt{4\pi D_\theta(t-\tau)}}\Bigg]
 =\cos{2\theta_0}e^{-4D_\theta(t-\tau)}\\
and &\int d\theta\frac{e^{-\frac{(\theta-\theta_0)^2}{4D_\theta(t-\tau)}}}{\sqrt{4\pi D_\theta(t-\tau)}}=1
\end{split}
\label{gauss_prop}
\end{equation}

Let us consider the trajectory of the particle over the interval $[\theta, t]$. If no resetting occurs during this interval, the position evolves according to the motion of standard active Brownian particle. For trajectories involving at least one resetting event, let the time elapsed since the last resetting be denoted by $\tau$. The position at time $t$ is then determined by the free ABP evolution during the interval $\tau$, starting from an arbitrary orientation $\theta_{t-\tau}$, which itself is governed by the Brownian motion of $\theta$. 

To compute the position distribution, we integrate over all possible values of $\tau \in [0, t]$ and $\theta_{t-\tau} \in [-\infty, \infty]$. Combining all these contributions, the renewal equation is expressed as:

\begin{equation}
    \begin{split}
        P(x,y,t)=e^{-rt}P_0(x,y,t)+r\int_{0}^{t}d\tau e^{-r\tau}\times\int_{-\infty}^{\infty}\mathbf{P}_0^{\theta}(x,y,\tau)\frac{e^{-\frac{\theta^2}{4D_\theta(t-\tau)}}}{\sqrt{4\pi D_\theta(t-\tau)}}
    \end{split}
\end{equation}
Here, we use the notation $\mathbf{P}_0^{\theta}(x, y, \tau)$ to represent the probability that the free active Brownian particle (ABP) is located at $(x, y)$ at time $\tau$, given that it started with an initial orientation $\theta$ at $\tau = 0$. The structure of the renewal equation described above differs significantly from that obtained for the position-orientation resetting protocol. Consequently, the resulting behavior is also anticipated to exhibit distinct characteristics.

The second moment for initial angle $\theta_0$ is calculated from Eq.(\ref{pos_reset_renewal}) and (\ref{gauss_prop}) as,

\begin{widetext}
\begin{equation}
\begin{split}
&\langle x^2(t)|\theta_0\rangle_{pr}=\frac{v_0^2}{r(r+D_\theta)}+\frac{2\Bar{D}}{r}+e^{-rt}\Bigg[\frac{v_0^2\cos{2\theta_0(r-6D_\theta)}}{6D_\theta^2(r-4D_\theta)}-\frac{v_0^2}{rD_\theta}-\frac{2\Bar{D}}{r}-\frac{\Delta D\cos{2\theta_0}}{r-4D_\theta}\Bigg]\\
&+e^{-4D_\theta t}\Bigg[\frac{v_0^2\cos{2\theta_0(r^2-6rD_\theta+12D_\theta^2)}}{12D_\theta^2(r-3D_\theta)(r-4D_\theta)}+\frac{\Delta D\cos{2\theta_0}}{r-4D_\theta}\Bigg]+e^{-(r+D_\theta)t}\Bigg[\frac{v_0^2}{D_\theta(r+D_\theta)}-\frac{v_0^2\cos{2\theta_0}(r-4D_\theta)}{4D_\theta^2(r-3D_\theta)}\Bigg]
\end{split}
\label{second_moment_pos}
\end{equation}
\end{widetext}

Putting $n=1$ in Eq.(\ref{1000}) we can find the value of $\langle x(t)|\theta_0\rangle_r$ for initial orientation $\theta_0=0$ as,

\begin{equation}
\langle x(t)|\theta_0\rangle_{pr}=\frac{v_0}{(r-D_\theta)}(e^{-D_\theta t}-e^{-rt})
\label{avg_x_pos}
\end{equation}

and $\langle y(t)\rangle=0$. Note that, for $r=D_\theta$ the above equation remains well defined when $t\ll D_\theta^{-1}$. At short times, i.e. for $t\ll min(r^{-1},D_\theta^{-1})$ 

\begin{equation}
    \langle x(t)\rangle_{pr}=v_0t-v_0(r+D_\theta)t^2+\mathcal{O}(t^3)
\end{equation}

The MSD along $x$ direction can be found as $\sigma_x^2(t)_{pr}=\langle x^2(t)|\theta_0=0\rangle_r-\langle x(t)|\theta_0=0\rangle_r^2$, here $\sigma_x^2(t)_{pr}$ denotes as MSD along $x$ due to position reset at $\theta_0=0$. 

\begin{widetext}
\begin{equation}
\begin{split}
\sigma_x^2(t)_{pr}&=\frac{v_0^2}{r(r+D_\theta)}+\frac{2\Bar{D}}{r}+e^{-rt}\Bigg[\frac{v_0^2(r-6D_\theta)}{6D_\theta^2(r-4D_\theta)}-\frac{v_0^2}{rD_\theta}-\frac{2\Bar{D}}{r}-\frac{\Delta D}{r-4D_\theta}\Bigg]+e^{-4D_\theta t}\Bigg[\frac{v_0^2(r^2-6rD_\theta+12D_\theta^2)}{12D_\theta^2(r-3D_\theta)(r-4D_\theta)}\\
&+\frac{\Delta D}{r-4D_\theta}\Bigg]+e^{-(r+D_\theta)t}\Bigg[\frac{v_0^2}{D_\theta(r+D_\theta)}-\frac{v_0^2(r-4D_\theta)}{4D_\theta^2(r-3D_\theta)}\Bigg]\-\frac{v_0^2}{(r-D_\theta)^2}(e^{-D_\theta t}-e^{-rt})^2
\end{split}
\label{msdx_pos}
\end{equation}
\end{widetext}

\begin{figure}
    \centering
    \includegraphics[width=1\textwidth]{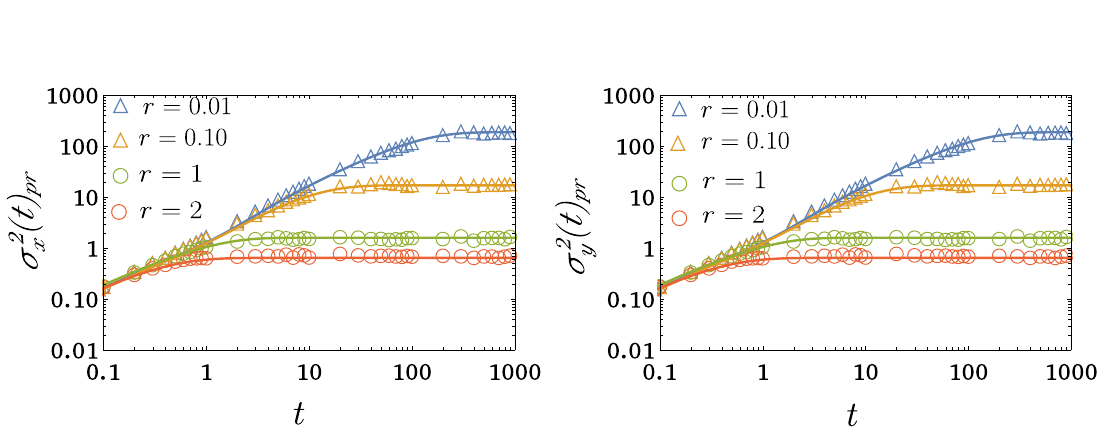}
    \caption{Position resetting: Mean square displacements along $x$ and $y$ as a function of time $t$ for $D_\theta=1$, $D_\parallel=1$, $D_\perp=0.1$, $v_0=1$, $\theta_0=0$ and different values of resetting rate $r$. Symbols represent the data from simulations. Solid lines are representing expressions from Eq.(\ref{msdx_pos}) and (\ref{msdy_pos}).}
    \label{fig5}
\end{figure}

Similarly MSD along $y$ axis becomes for $\theta_0=0$

\begin{widetext}
\begin{equation}
\begin{split}
\sigma_y^2(t)_{pr}&=\frac{v_0^2}{r(r+D_\theta)}+\frac{2\Bar{D}}{r}-e^{-rt}\Bigg[\frac{v_0^2(r-6D_\theta)}{6D_\theta^2(r-4D_\theta)}+\frac{v_0^2}{rD_\theta}+\frac{2\Bar{D}}{r}-\frac{\Delta D}{r-4D_\theta}\Bigg]\\
&+e^{-4D_\theta t}\Bigg[-\frac{\Delta D}{r-4D_\theta}-\frac{v_0^2(r^2-6rD_\theta+12D_\theta^2)}{12D_\theta^2(r-3D_\theta)(r-4D_\theta)}\Bigg]+e^{-(r+D_\theta)t}\Bigg[\frac{v_0^2}{D_\theta(r+D_\theta)}+\frac{v_0^2(r-4D_\theta)}{4D_\theta^2(r-3D_\theta)}\Bigg]
\end{split}
\label{msdy_pos}
\end{equation}
\end{widetext}

If we expand the exponential terms of Eqs.(\ref{msdx_pos}) and (\ref{msdy_pos}) up to higher order, we obtain

\begin{widetext}
\begin{equation}
\begin{split}
\sigma_x^2(t)_{pr}&=\Big(2D_{\parallel}+\frac{rv_0^2}{12D_\theta^2}\Big)t+\Bigg(-\frac{v_0^2}{2}-r\Bar{D}-\frac{\Delta D(r+4D_\theta)}{2}+\frac{v_0^2}{3(3D_\theta-r)(4D_\theta-r)}\Big(18D_\theta^2\\
&-21rD_\theta+\frac{49r^2}{8}-\frac{r^4}{8D_\theta^2}+\frac{9r^3}{4D_\theta}-\frac{9r^3}{4}\Big)\Bigg)t^2+\Bigg(\frac{v_0^2(r+D_\theta)(24D_\theta^2-4rD_\theta-r^2)}{8D_\theta(3D_\theta-r)}-\frac{rv_0^2}{2}\Bigg)t^3+\mathcal{O}(t^4)
\end{split}
\label{posshort}
\end{equation}

\begin{equation}
\begin{split}
\sigma_y^2(t)_{pr}&=\frac{v_0^2(r-4D_\theta)}{4D_\theta(r-3D_\theta)(r-4D_\theta)}+\Bigg(2D_\perp+\frac{v_0^2(27r^2D_\theta-12rD_\theta^2-r^3-48D_\theta^3)}{12D_\theta^2(r-3D_\theta)(r-4D_\theta)}+\frac{v_0^2(r^2-6rD_\theta+12D_\theta^2)}{3D_\theta(r-3D_\theta)(r-4D_\theta)}\Bigg)t\\
&+\Bigg(\frac{\Delta D(r+4D_\theta)}{2}-r\Bar{D}-\frac{v_0^2r}{2D_\theta}-\frac{r^2v_0^2(r-6D_\theta)}{12D_\theta^2(r-4D_\theta)}-\frac{2v_0^2(r^2-6rD_\theta+12D_\theta^2)}{(r-4D_\theta)(r-3D_\theta)}\\
&+\frac{v_0^2(r-D_\theta)^2(r^2+rD_\theta-16D_\theta^2)}{8D_\theta^2(r-3D_\theta)(r+D_\theta)}\Bigg)t^2-\frac{rv_0^2}{2}\Bigg(1+\frac{(r+D_\theta)(r-4D_\theta)}{4D_\theta(r-3D_\theta)}\Bigg)t^3+\mathcal{O}(t^4)
\end{split}
\label{posyshort}
\end{equation}
\end{widetext}

At short times, Eqs.(\ref{posshort}) and (\ref{posyshort}) indicate that the system exhibits super-diffusive behavior, showing the anisotropy in the dynamics. Conversely, at long times $t \gg \max(r^{-1}, D_\theta^{-1})$, both $\sigma_x^2$ and $\sigma_y^2$ settle into a stationary state. Interestingly, at this stage, their values become identical.

\begin{equation}
    \begin{split}
        (\sigma_x^2)_{pr}=(\sigma_y^2)_{pr}=\frac{v_0^2}{r(r+D_\theta)}+\frac{2\Bar{D}}{r}
    \end{split}
    \label{stationary_pos}
\end{equation}

Eq.(\ref{stationary_pos}) demonstrates that the anisotropy disappears in the stationary state, resulting in effective values for the mean squared displacements (MSDs) that are influenced by the propulsion velocity $v_0$ of the particle and the average diffusion coefficient $\Bar{D}$. Fig.(\ref{fig5}) presents the plots of $\sigma_x^2(t)_{pr}$ and $\sigma_y^2(t)_{pr}$ for various resetting rates, along with the corresponding results from numerical simulations. The leading order time-scale of relaxation to the stationary value is $\tau_{relax}^{pos}=\min[r^{-1}, (4D_\theta)^{-1}, (r+D_\theta)^{-1}]$. There is a very interesting difference in relaxation time-scale of the first and second type resetting protocols. In the position-only reset, angular diffusion ($4D_\theta$ term) plays a significant independent role since orientation dynamics are not reset. When rotational diffusion constant is very very small then the relaxation time scale depends on the term $4D_\theta$. This adds a separate time scale to the competition. In the position-and-orientation reset, the dynamics of orientation are directly tied to the reset rate $r$, reducing the competition to only two combined rates. For position-only reset, when $D_\theta$ is very small, the relaxation time scale depends on $4D_\theta$. For position-and-orientation reset, when $D_\theta$ is very small, the relaxation time scale depends on the reset rate $r$, as orientation is reset along with position.

\begin{figure}
    \centering
    \includegraphics[width=1\linewidth]{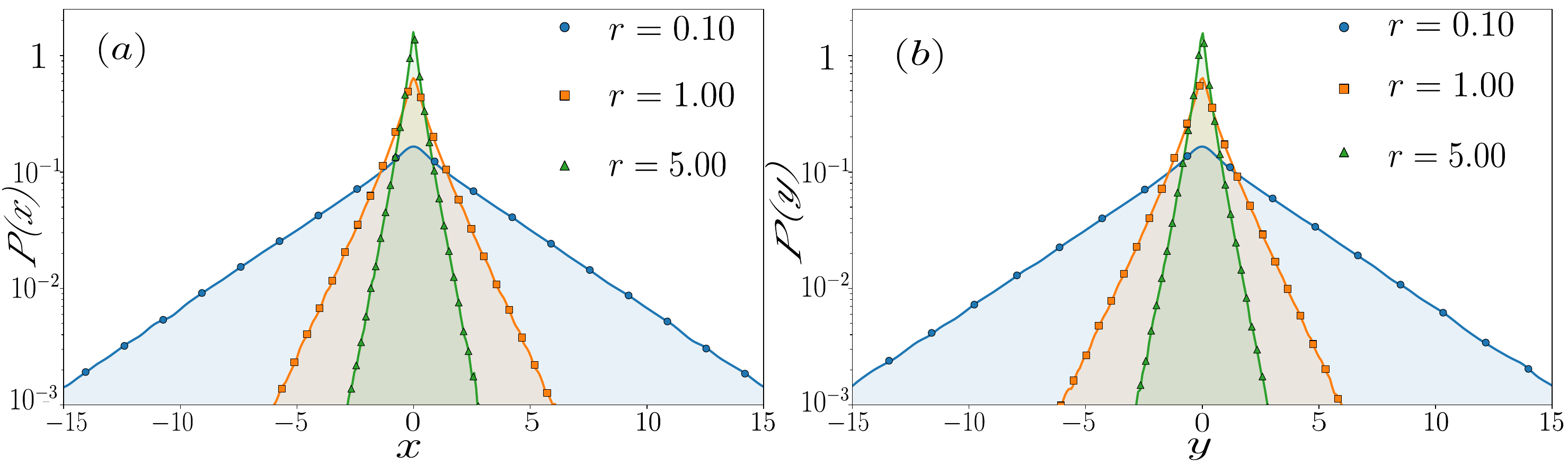}
    \caption{Position distribution along $x$ and $y$ of anisotropic ABP under position resetting at $t=100$ (stationary state) when $D_\theta=0.1$, $D_\parallel=1$, $D_\perp=0.1$, $v_0=1$ and $\theta_0=0$ for different values of reset rate $r$ as shown in the legends.}
    \label{positionx}
\end{figure}

Figs.(\ref{positionx}(a)) and (\ref{positionx}(b)) present the simulation results for the position distribution in the long-time stationary state for both the $x$ and $y$ components. As observed in the previous section, the $x$ and $y$ position probability distributions are distinct under position-orientation resetting. However, in the case where only position resetting is implemented, the probability distributions become identical in both the $x$ and $y$ directions. When position resetting occurs, the particle's position is reset independent of its orientation or velocity direction. The subsequent motion, influenced by the anisotropy, may lead to an anisotropic trajectory, but the distribution of positions after the reset is governed by the reset distribution itself, which is usually symmetric in both the $x$ and $y$ directions. Even if the particle is biased in a particular direction by its self-propulsion, the reset process essentially "erases" the effect of previous motion, resetting the position in a symmetric manner.

\subsection{Orientational Resetting}\label{angle_reset}
Let us now study the case of orientational resetting, when $\theta$ undergoes reset to $\theta=0$ with rate $r$, while $x$ and $y$ diffuse usually. In this scenario, the position distribution does not directly adhere to any renewal equation, whereas the $\theta$ distribution does. We follow the general discussion of the renewal equation approach

\begin{equation}
\begin{split}
&P_r(x,y,\theta,t|0,0,\theta_0)=e^{-rt}P_0(x,y,\theta,t|0,0,\theta_0)+r\int_{0}^{t}d\tau e^{-r\tau}\\
&\int dx^\prime\int dy^\prime\int d\theta^\prime P_r(x^\prime,y^\prime,\theta^\prime,t-\tau|0,0,\theta_0)P_0(x,y,\theta,\tau|x^\prime,y^\prime,\theta_0)
\end{split}
\end{equation}
We examine the third resetting protocol, in which the orientation \(\theta\) resets to \(\theta = 0\) at a rate \(r\), while the position remains unchanged. Let \(P(\theta, t | \theta^\prime, t^\prime)\) represent the probability that the orientation is \(\theta\) at time \(t\), given it was \(\theta^\prime\) at an earlier time \(t^\prime\). This conditional probability \(P(\theta, t | \theta^\prime, t^\prime)\) adheres to a renewal equation:
\begin{equation}
\begin{split}
P(\theta,t|\theta^\prime,t^\prime)=e^{-r(t-t^\prime)}P_0(\theta,t|\theta^\prime,t^\prime)+r\int_{0}^{t-t^\prime}d\tau e^{-r\tau}P_0(\theta,\tau|0,0)
\end{split}
\end{equation}
where $P_0(\theta,t|\theta^\prime,t^\prime)$ denotes the propagator for the standard Brownian motion.

The final form of the second moment of $x$ can be found using Eqs.(\ref{B7}),(\ref{B9}), and (\ref{B13}) (detailed calculations are done in Appendix(\ref{appC})).

\begin{widetext}
\begin{equation}
\begin{split}
&\langle x^2(t)|\theta_0\rangle_{or}=\Bigg[\Big(2\Bar{D}+\frac{r\Delta D\cos{2\theta_0}}{r+4D_\theta}\Big)t+\frac{4D_\theta\Delta D\cos{2\theta_0}}{(r+4D_\theta)^2}(1-e^{-(r+4D_\theta)t})\Bigg]+\frac{v_0^2}{(r+D_\theta)^2}\Bigg[r^2t^2\\
&+\frac{2D_\theta t(2D_\theta^2+9rD_\theta+r^2)}{(r+D_\theta)(r+4D_\theta)}\Bigg]+\frac{4v_0^2e^{-(r+4D_\theta)t}}{3(r+4D_\theta)^2}-\frac{6v_0^2D_\theta^2(2D_\theta^2+16rD_\theta+5r^2)}{(r+D_\theta)^4(r+4D_\theta)^2}+\frac{2v_0^2e^{-(r+D_\theta)t}}{(r+D_\theta)^3}\Bigg(rD_\theta t\\
&+\frac{4D_\theta^3+33rD_\theta^2-2r^3}{3(r+D_\theta)(r+4D_\theta)}\Bigg)
\end{split}
\end{equation}
\end{widetext}

At $\theta_0=0$ the above expression becomes,

\begin{widetext}
\begin{equation}
\begin{split}
&\langle x^2(t)|\theta_0=0\rangle_{or}=\Bigg[\Big(2\Bar{D}+\frac{r\Delta D}{r+4D_\theta}\Big)t+\frac{4D_\theta\Delta D}{(r+4D_\theta)^2}(1-e^{-(r+4D_\theta)t})\Bigg]+\frac{v_0^2}{(r+D_\theta)^2}\Bigg[r^2t^2+\frac{2D_\theta t(2D_\theta^2+9rD_\theta+r^2)}{(r+D_\theta)(r+4D_\theta)}\Bigg]\\
&+\frac{4v_0^2e^{-(r+4D_\theta)t}}{3(r+4D_\theta)^2}-\frac{6v_0^2D_\theta^2(2D_\theta^2+16rD_\theta+5r^2)}{(r+D_\theta)^4(r+4D_\theta)^2}+\frac{2v_0^2e^{-(r+D_\theta)t}}{(r+D_\theta)^3}\Bigg(rD_\theta t+\frac{4D_\theta^3+33rD_\theta^2-2r^3}{3(r+D_\theta)(r+4D_\theta)}\Bigg)
\end{split}
\label{43}
\end{equation}
\end{widetext}

The MSD $\sigma_x^2(t)_{or}=\langle x^2(t)|\theta_0\rangle_{or}-\langle x(t)|\theta_0\rangle_{or}^2$ is then given by using Eqs.(\ref{43}) and (\ref{7})

\begin{widetext}
\begin{equation}
\begin{split}
& \sigma_x^2(t)_{or}=\Bigg[\Big(2\Bar{D}+\frac{r\Delta D}{r+4D_\theta}\Big)t+\frac{4D_\theta\Delta D}{(r+4D_\theta)^2}(1-e^{-(r+4D_\theta)t})\Bigg]+\frac{v_0^2}{(r+D_\theta)^2}\Bigg[r^2t^2+\frac{2D_\theta t(2D_\theta^2+9rD_\theta+r^2)}{(r+D_\theta)(r+4D_\theta)}\Bigg]\\
&+\frac{4v_0^2e^{-(r+4D_\theta)t}}{3(r+4D_\theta)^2}-\frac{6v_0^2D_\theta^2(2D_\theta^2+16rD_\theta+5r^2)}{(r+D_\theta)^4(r+4D_\theta)^2}+\frac{2v_0^2e^{-(r+D_\theta)t}}{(r+D_\theta)^3}\Bigg(rD_\theta t+\frac{4D_\theta^3+33rD_\theta^2-2r^3}{3(r+D_\theta)(r+4D_\theta)}\Bigg)\\
&-\frac{v_0^2}{(r+D_\theta)^2}\Bigg[rt+\frac{D_\theta}{r+D_\theta}(1-e^{-(r+D_\theta)t})\Bigg]^2
\end{split}
\label{msdx_orientation}
\end{equation}
\end{widetext}

\begin{figure}
    \centering
    \includegraphics[width=1\linewidth]{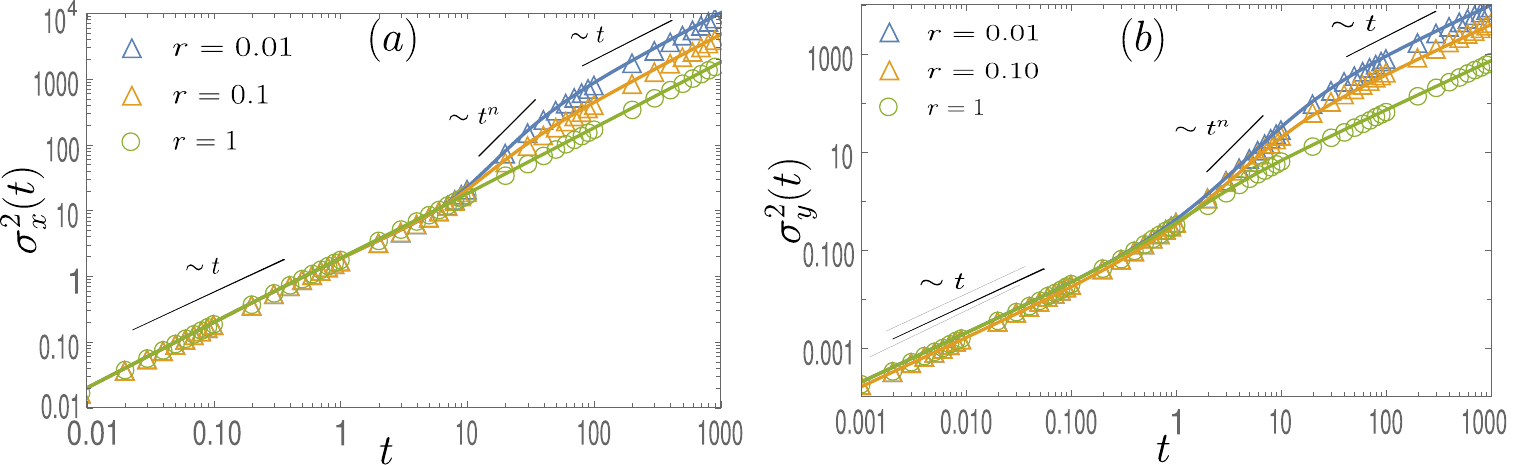}
    \caption{Orientation resetting: Mean square displacements along $x$ and $y$ directions as a function of time $t$ for $D_\theta=0.1$, $D_\parallel=1$, $D_\perp=0.1$, $v_0=1$, $\theta_0=0$ and varying reset rate $r$. Symbols represent the data from simulations. Solid lines are representing expression from Eqs.(\ref{msdx_orientation}) and (\ref{msdy_orientation}).}
    \label{msdx_angle}
\end{figure}

Similarly, we can calculate $\langle y^2(t)|\theta_0=0\rangle_{or}$ in Eq.(\ref{C15}) and as the average of $y$ at $\theta_0=0$ then the second moment of $y$ and MSD along $y$ both are same as expressed

\begin{widetext}
\begin{equation}
    \begin{split}
        \sigma_y^2(t)_{or}&=\langle y^2(t)|\theta_0=0\rangle=\Bigg[\Bigg(2\bar{D}-\frac{r\Delta D}{r+4D_\theta}\Bigg)t-\frac{4D_\theta\Delta D}{(r+4D_\theta)^2}\Big(1-e^{-(r+4D_\theta)t}\Big)\Bigg]+\frac{4v_0^2D_\theta t}{(r+4D_\theta)(r+D_\theta)}\\
       &-\frac{4v_0^2D_\theta(2r+5D_\theta)}{(r+4D_\theta)^2(r+D_\theta)^2}+\frac{4v_0^2}{3}\Bigg[\frac{e^{-(r+D_\theta)t}}{(r+D_\theta)^2}-\frac{e^{-(r+4D_\theta)t}}{(r+4D_\theta)^2}\Bigg]
    \end{split}
    \label{msdy_orientation}
\end{equation}
\end{widetext}

Fig. (\ref{msdx_angle}(a)) and (\ref{msdx_angle}(b)) present the simulation results of the mean-square displacement (MSD) along the $x$ and $y$-axes respectively as a function of time $t$. These results align well with the analytical expressions derived in Eq.(\ref{msdx_orientation}) and (\ref{msdy_orientation}), validating the theoretical framework. These figures illustrate that the particle's dynamics do not reach a steady state despite the implementation of orientational resetting. This behavior contrasts with scenarios involving complete or solely positional resetting protocols, which typically drive the system toward steady-state dynamics. The dynamics initially exhibit pure diffusive behavior, transitioning into a super-diffusive regime over intermediate times. At long times, the system eventually reverts to a diffusive nature. An intriguing phenomenon observed in this case is the persistence of anisotropy even in the long-time regime, as evidenced by both the simulation results and the analytical expression in Eq.(\ref{diff_orient_reset}). This contrasts sharply with the scenario under no-reset conditions, as discussed in Mandal et al. \cite{mandal2024diffusion}, where the system transitions to isotropic behavior at long time. This highlights the role of orientational resetting in maintaining anisotropic dynamics over extended periods.
In the intermediate super-diffusive regime, characterized by a time dependence of $t^n$ (where $n>1$), the degree of super-diffusivity is strongly influenced by the resetting rate $r$. Lower resetting rates amplify the super-diffusive behavior, making it more pronounced, whereas higher resetting rates tend to suppress it, steering the dynamics more quickly toward diffusive behavior. The results corresponding to only orientational resetting present a stark contrast to those reported by Kumar et al. \cite{kumar2020active}. In their study, the isotropic nature of active Brownian particles (ABPs) precluded any initial diffusive behavior. Here, however, the introduction of anisotropy leads to an initial diffusive regime, highlighting the significant impact of anisotropic properties on the dynamics of the system. To better understand the nature of these crossovers, we investigate the short-term and long-term behaviors of the mean square displacements.

\begin{figure}
    \centering
\includegraphics[width=0.8\linewidth]{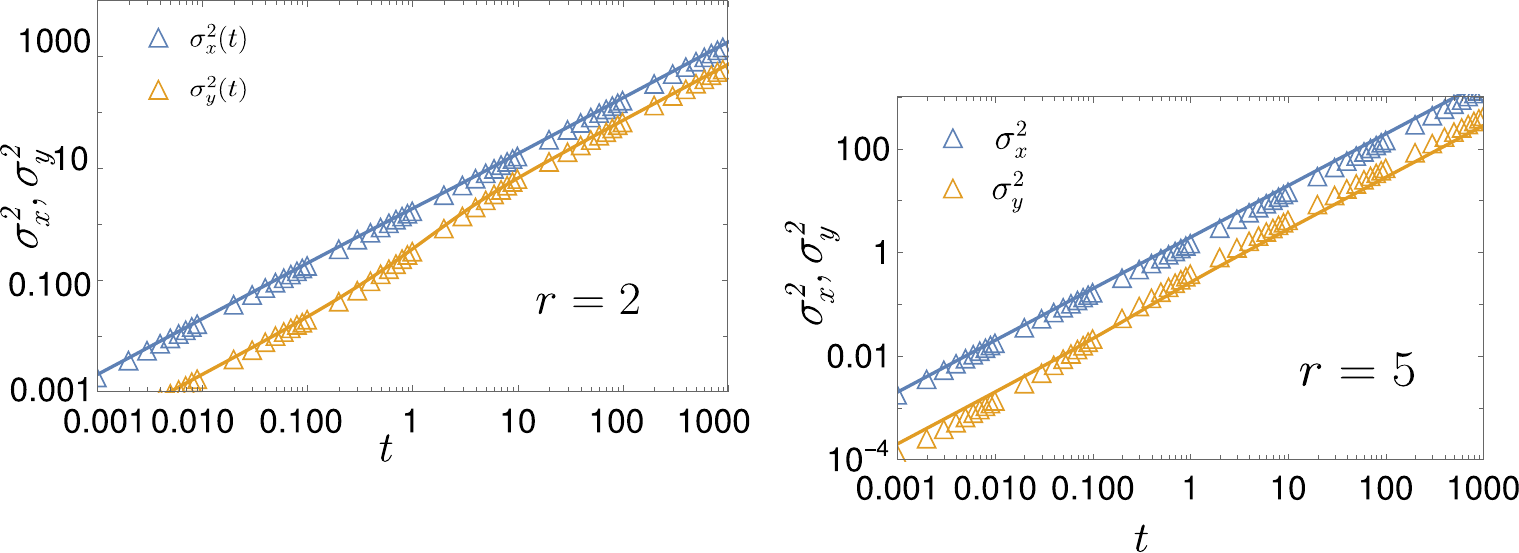}
\caption{Orientation resetting: Comparison of mean-square displacements along $x$ and $y$- axes for $D_\theta=0.1$, $D_\parallel=1$, $D_\perp=0.1$, $v_0=1$, $\theta_0=0$. Two sets of reset rate $r=1$, and $r=5$ have been taken. Symbols represent the data from simulations. Solid lines are representing expression from Eqs.(\ref{msdx_orientation}) and (\ref{msdy_orientation}).}
    \label{msdxy_angle}
\end{figure}

\begin{figure}
    \centering
\includegraphics[width=1\linewidth]{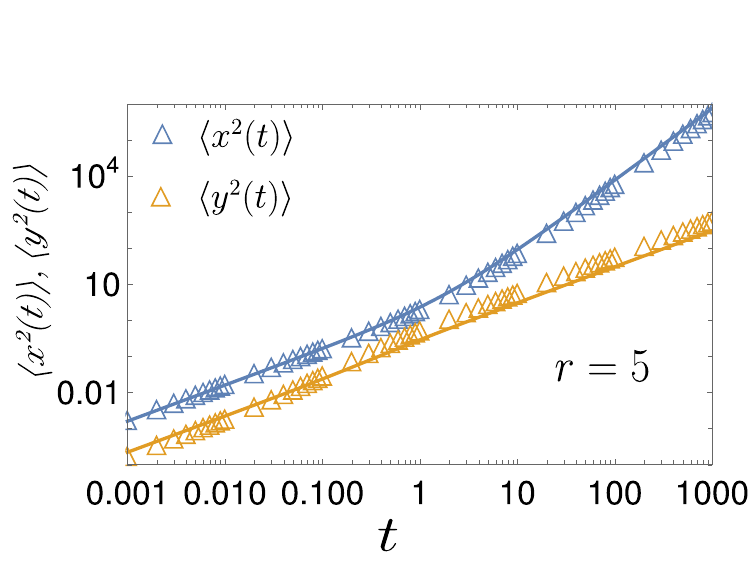}
\caption{Orientation resetting: Comparison of second moments along $x$ and $y$- axes for $D_\theta=0.1$,$D_\parallel=1$, $D_\perp=0.1$, $v_0=1$, $\theta_0=0$ and $r=5$. Symbols represent the data from simulations. Solid lines are representing expression from Eqs.(\ref{43}) and (\ref{msdy_orientation}).}
    \label{second_moment_xy_angle}
\end{figure}

At very short-times, we obtain the MSD $x$ and $y$- components by expanding the exponential terms 

\begin{widetext}
\begin{equation}
    \begin{split}
        \sigma_x^2(t)_{or}&=\frac{4v_0^2}{3(r+4D_\theta)^2}-\frac{v_0^2D_\theta^2}{(r+D_\theta)^4}+\Bigg[2D_{\parallel}+\frac{20v_0^2D_\theta^3-4v_0^2r^3+12r^2D_\theta(1-v_0^2)+6rD_\theta^2(8-v_0^2)}{3(r+D_\theta)^3(r+4D_\theta)}\Bigg]t\\
        &+\Bigg[\frac{2v_0^2}{3}-2D_\theta\Delta D-\frac{2v_0^2(r^3+4D_\theta^3+9rD_\theta^2+6r^2D_\theta)}{3(r+D_\theta)^2(r+4D_\theta)}\Big]t^2+\Bigg[\frac{2rv_0^2D_\theta(10D_\theta^3-3r^3-24rD_\theta^2-12r^2D_\theta)}{3(r+D_\theta)^3(r+4D_\theta)}\\
        &+\frac{8r\Delta DD_\theta^2}{r+4D_\theta}+\frac{2rv_0^2D_\theta}{r+D_\theta}\Bigg]t^3+\Bigg[\frac{16r^2D_\theta^2v_0^2}{3(r+4D_\theta)^2}-\frac{16r^2D_\theta^3\Delta D}{(r+4D_\theta)^2}-\frac{4r^2v_0^2D_\theta^4}{(r+D_\theta)^4}-\frac{2r^2D_\theta^2 v_0^2}{(r+D_\theta)^2}\\
       &+\frac{r^2v_0^2D_\theta^2(8D_\theta^3+18rD_\theta^2-r^3)}{3(r+D_\theta)^4(r+4D_\theta)}\Bigg]t^4+\mathcal{O}(t^5)
    \end{split}
\end{equation}

\begin{equation}
    \begin{split}
        \sigma_y^2(t)_{or}&=\Bigg(2D_{\perp}-\frac{4D_\theta v_0^2}{3(r+D_\theta)(r+4D_\theta)}\Bigg)t+2D_\theta\Delta D t^2+\Bigg[\frac{8v_0^2rD_\theta}{3(r+4D_\theta)}-\frac{2rD_\theta v_0^2}{3(r+D_\theta)}-\frac{8r\Delta DD_\theta^2(r+4)}{(r+4D_\theta)^2}\Bigg]t^3\\
        &+\Bigg[\frac{r^2D_\theta^2v_0^2}{3(r+D_\theta)^2}+\frac{16r^2D_\theta^2(\Delta D+v_0^2/3)}{(r+4D_\theta)^2}\Bigg]t^4+\mathcal{O}(t^5)
    \end{split}
\end{equation}
\end{widetext}

The transient behavior can once again be interpreted through an analysis of the dynamical exponents governing the motion along the $x$ and $y$ directions. In the long-term limit, since all terms increase linearly over time, we can ignore the constant offset in the equations above, and the MSD along the $x$ and $y$ axes are given by,
\begin{equation}
    \begin{split}
       & \sigma_x^2(t)_{or}\approx\Big(2\bar{D}+\frac{r\Delta D}{r+4D_\theta}+\frac{2v_0^2D_\theta^2(2D_\theta+5r)}{(r+D_\theta)^3(r+4D_\theta)}\Big)t\\
         &\sigma_y^2(t)_{or}\approx\Bigg(2\bar{D}-\frac{r\Delta D}{r+4D_\theta}+\frac{4v_0^2D_\theta}{(r+4D_\theta)(r+D_\theta)}\Bigg)t
    \end{split}
\end{equation}
The dynamics in $x$ and $y$ directions conversely goes through super-diffusive regime, before also re-entering a diffusive regime. One of the notable observations from the second moments of both $x$ and $y$ 
is that $\langle x^2(t) \,|\, \theta_0 = 0 \rangle_{\text{or}}$ exhibits 
super-diffusive behavior found in Eq.(\ref{43}), scaling as $t^2$ even in the long-time limit, 
whereas $\langle y^2(t) \,|\, \theta_0 = 0 \rangle_{\text{or}}$ remains 
diffusive at long times found in Eq.(\ref{msdy_orientation})(see fig(\ref{second_moment_xy_angle})). Although the second moments along the two axes 
differ in nature, the mean-squared displacement (MSD) along both axes 
shows diffusive behavior in the long-time limit.

Fig.(\ref{msdxy_angle}) presents a comparative analysis of the mean-square displacements (MSD) along the $x$ and $y$ directions for a specific reset rates $r=1$ and $r=5$. The figures reveal that the anisotropy observed at early times persists even at long times due to the orientational resetting. This behavior is particularly intriguing when contrasted with the no-reset condition, as discussed in the work by Mandal et al.\cite{mandal2024diffusion}. For the no-reset condition, the initial anisotropy observed in the anisotropic ABP dynamics diminishes over time, eventually vanishing at long times, leading the particle to behave like an isotropic one.
 The effect of resetting becomes apparent at higher orders, introducing an additional anisotropy. This behavior is expected, as the resetting condition $\theta=0$ is fundamentally anisotropic in nature. Interestingly, the impact of this anisotropy persists even at late times despite the fact that both \(x\)- and \(y\)-motions ultimately become diffusive. The long-time anisotropy becomes more pronounced with an increase in the reset rate.

In the long-term limit, we can anticipate that according to the central limit theorem, the two-dimensional probability distribution in the Cartesian components takes on a Gaussian form, and is given by

\begin{equation}
    P(x,y,t)=\frac{1}{4\pi\sqrt{Det(\mathbf{D}_{eff})}}exp\Bigg[-\frac{1}{4t}(x,y)^T\mathbf{D}^{-1}_{eff}(x,y)\Bigg]
\end{equation}
where the effective diffusion coefficients $\mathbf{D}$ can be written for $x$ and $y$ axes as,
\begin{equation}
    \begin{split}
        &D_{xx}^{eff}\approx 2\bar{D}+\frac{r\Delta D}{r+4D_\theta}+\frac{2v_0^2D_\theta^2(2D_\theta+5r)}{(r+D_\theta)^3(r+4D_\theta)}\\
        &D_{yy}^{eff}\approx 2\bar{D}-\frac{r\Delta D}{r+4D_\theta}+\frac{4v_0^2D_\theta}{(r+4D_\theta)(r+D_\theta)}
    \end{split}
    \label{diff_orient_reset}
\end{equation}
The marginal probability distribution of $x$ can be found by integrating out the $y$ variable, as given by

\begin{equation}
    P(x,t)=\int dyP(x,y,t)=\frac{1}{\sqrt{4\pi D_{eff}^xt}}exp\Big[-\frac{(x-\mu_x(t))^2}{4D_{eff}^xt}\Big]
\end{equation}
Where $\mu_x(t)=\langle x(t)\rangle$ which has finite value as shown in the Eq.(\ref{7}). A similar argument applies to the marginal probability distribution of $y$, where the mean $\mu_y(t)$ is always zero. One of the most intriguing outcomes of applying orientation resetting in anisotropic ABP is the absence of a stationary state over extended time periods, with the anisotropy persisting. However, in the case of ABP without resetting, the anisotropy gradually diminishes in the long-time limit\cite{mandal2024diffusion}.

\begin{figure}
\centering
    \includegraphics[width=1\linewidth]{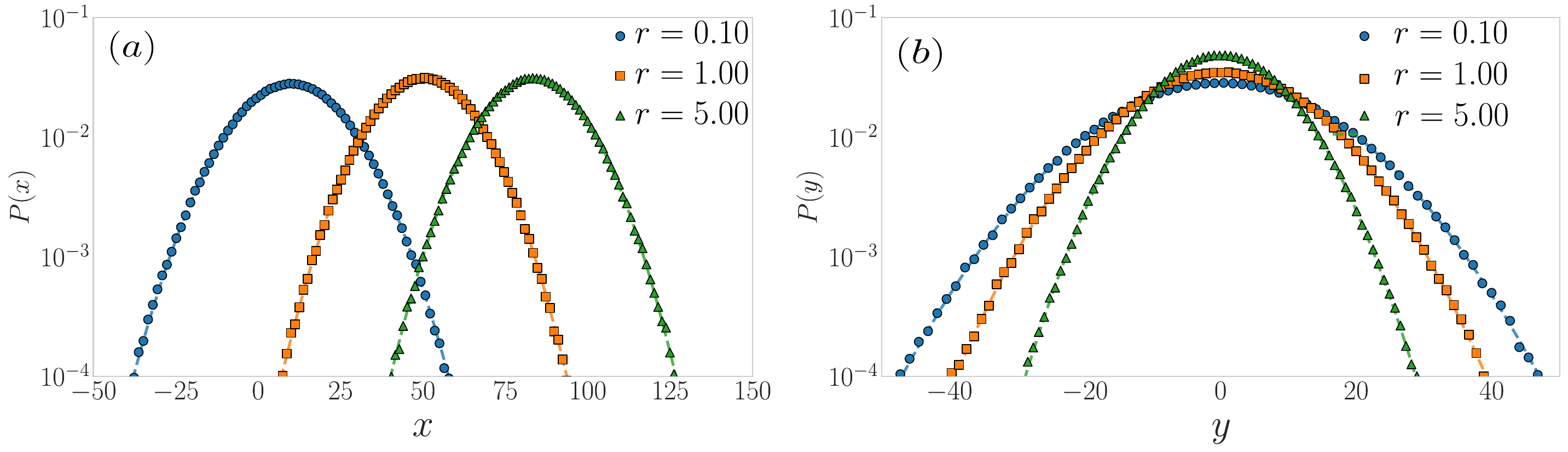}
    \caption{Plot of long time position distribution along $x$ and $y$ directions of anisotropic ABP under orientation resetting when $D_\theta=0.1$, $D_\parallel=1$, $D_\perp=0.1$, $v_0=1$ and $\theta_0=0$ at $t=100$ for different values of reset rate $r$ as shown in the legends.}
    \label{orient_prob_x}
\end{figure}

Figs.(\ref{orient_prob_x}(a)) and (\ref{orient_prob_x}(b)) depict the simulation results of position distributions along the $x$- and $y$-axes, respectively. These distributions exhibit Gaussian profiles, reflecting the expected behavior of the particle under the influence of orientational resetting. This Gaussian nature underscores the underlying stochastic processes governing the particle's dynamics in both spatial dimensions. The propulsion velocity, aligned with the major axis of the particle, causes the orientation resetting to shift the Gaussian peak in the positive $x$-direction as the resetting rate increases. This directional bias emerges because higher resetting rates reinforce alignment along the propulsion axis, effectively steering the particle's position distribution towards the positive $x$-axis over time.

\begin{figure}
    \centering
\includegraphics[width=1\linewidth]{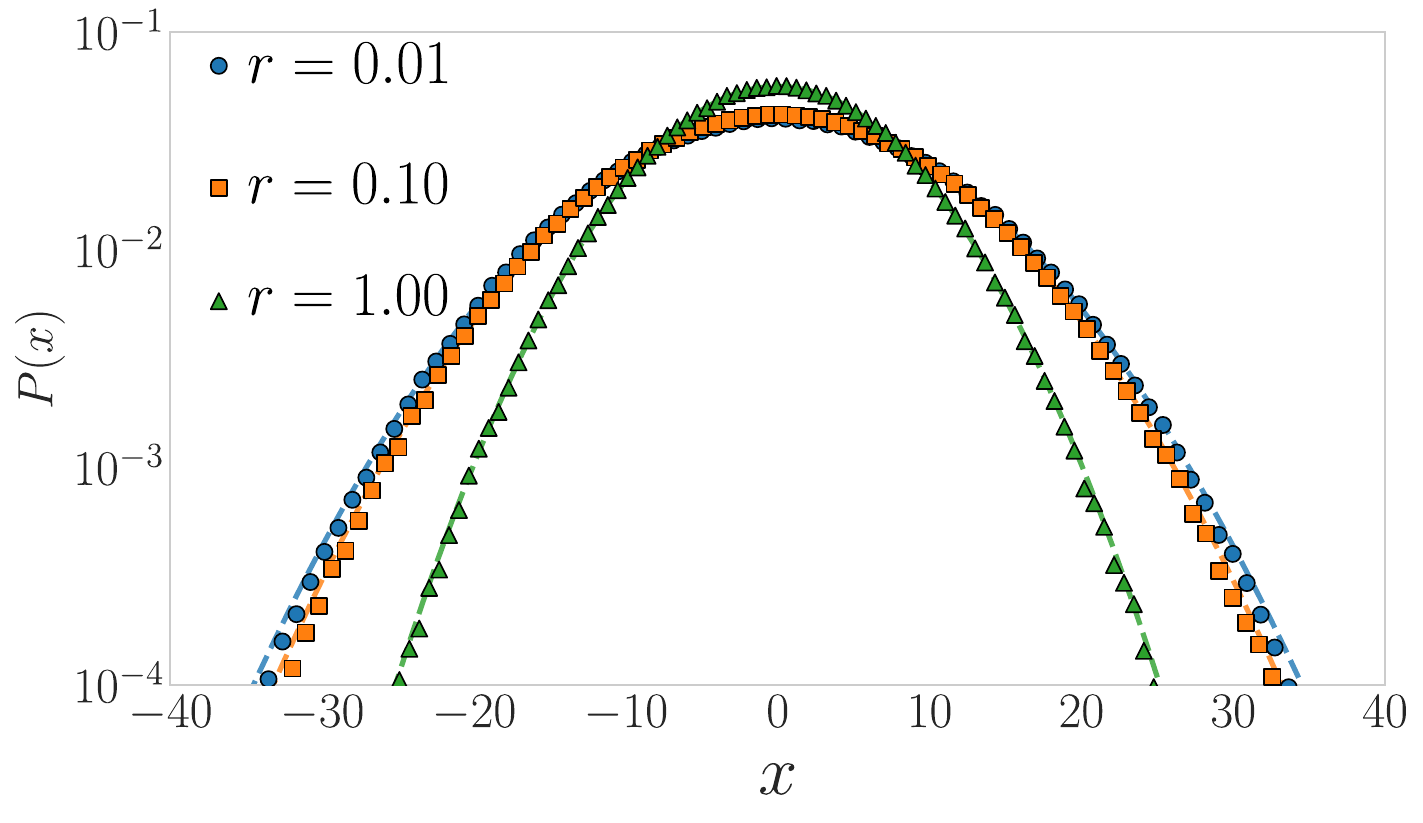}
\caption{$4$-state orientation resetting: Mapping with RTP along $x$ axis for $D_\theta=0.1$, $D_\parallel=1$, $D_\perp=0.1$, $v_0=1$ at $t=100$. Different reset rates are mentioned in the legends.}
    \label{4-state}
\end{figure}

 One interesting property of orientation resetting in an anisotropic ABP is that it can be mapped to a run-and-tumble(RTP)–like particle when multiple orientation reset angles are considered. There can be multiple orientation states, ranging from $2$ states up to $n$ states. In the $2$-state RTP, the anisotropic ABP can be considered to reset to two distinct orientation angles, $\theta_0 = 0$ and $\pi$, respectively, following a Poissonian distribution. In the case of $3$-state orientation resetting with reset angles $\theta_0 = 0, \tfrac{2\pi}{3},$ and $\tfrac{4\pi}{3}$, the particle dynamics can be mapped to that of a $3$-state RTP. This model can be extended to an $n$-state RTP, where $n$ may represent either a finite or an infinite number of states. Such a type of RTP model in two dimensions has been discussed in the work of Santra \textit{et al.}~\cite{santra2020ru}. In $n$-state orientational resetting, the particle exhibits Gaussian distributions along both the $x$ and $y$ directions. However, the major effect is observed in the $x$-direction, where the distribution is again Gaussian with zero mean, in contrast to the single-state orientation reset along $\theta_0 = 0$. A simulation result for 4-state resetting is shown in Fig.~(\ref{4-state}), whose characteristics closely match those of the 4-state RTP in two dimensions studied by Santra \textit{et al.}~\cite{santra2020ru} at long time. The reset orientation angles are chosen as $0$, $\tfrac{\pi}{2}$, $\pi$, and $\tfrac{3\pi}{2}$. Upon resetting, the particle selects any of these four orientation states with equal probability. The simulation results suggest that an anisotropic ABP with multiple reset angles can be effectively mapped onto the run-and-tumble dynamics of a particle in a two-dimensional plane.

\begin{widetext}

\section{Conclusions} \label{sec:conclusions}

In this article, we investigated how shape asymmetry influences the two-dimensional ABP of an anisotropic particle under stochastic resetting. Specifically, we examined the behavior of moments, and steady-state distributions at both short and long times, considering different resetting schemes. It is well-established that anisotropic diffusion only persists over very short time scales, eventually transitioning to isotropic diffusion at longer times. In practical scenarios, the short-time dynamics are often insignificant, enabling the motion of an asymmetric particle to be effectively modeled by Langevin equations for a point particle. This simplification uses an isotropic diffusion coefficient, calculated as the mean of the diffusion coefficients along the particle's principal and secondary axes. This approximation has been successfully applied in contexts such as free particle motion or when the particle is confined harmonically in two dimensions \cite{grima2007brownian}.
For the case of complete resetting, we have derived both the time-dependent moments and a perturbative steady-state solution. Our analysis includes exact calculations up to the first-order approximation of anisotropy, providing a detailed understanding of the system's behavior under resetting conditions. We have demonstrated that the orientation-only resetting scheme is insufficient to drive the system to a steady state. However, higher resetting rates effectively suppress the intermediate hyper-diffusivity observed in the dynamics of the anisotropic ABP. Conversely, when orientational resetting is introduced, the short-time anisotropy in the particle's motion persists into late times, reflecting the influence of resetting on the dynamics.

This study paves the way for extending investigations into dynamical first passage properties, counting statistics, exit time statistics, and cover time for anisotropic biological particles, which are predominantly observed in nature.
\end{widetext}

\section{Acknowledgment}
We thank Prof. Sanjib Sabhapandit of RRI, Bengaluru, for his valuable comments. AG is grateful to RRI, Bengaluru, for access to their computing facilities and acknowledges Prof. Dipanjan Chakraborty of IISER Mohali for his support. SC acknowledges Prof. Hartmut Löwen for insightful discussions.

\section{Conflict of Interest}
The authors have no conflicts to disclose.

	\newpage    
\bibliographystyle{unsrt}
\bibliography{reference}	

\newpage   

\newpage 
	    
\onecolumngrid
\begin{center}
{\bf {\Large{Appendix}}}
\end{center}
\begin{appendix}
\section{Perturbative steady state for weak-asymmetric particles}\label{perturbative}
Although we obtain exact formulas for the lower-order moments, we are also interested in gaining a deeper understanding of the overall form of the non-equilibrium steady state (NESS) under resetting. For this, we adopt a simplified perturbative method. First, we compute the propagator without considering resetting, and then we apply the renewal framework to incorporate resetting and determine the steady state.

Without resetting, the dynamics of the system are governed by the Smoluchowski-Perrin equation, which provides a framework for describing the motion of anisotropic particles in a diffusive medium.
\begin{equation}
    \partial_tP_0(x,t|x_0)=D_\theta\partial^2_\theta P_0(x,t|x_0)+\nabla.[\mathbf{D}.\nabla]P_0(x,t|x_0)-v\hat{n}.\nabla P_0(x,t|x_0)
\end{equation}

where $\mathbf{D}$ is the diffusion tensor, $\mathbf{D}_{ij}=k_BT\Gamma_{ij}=D_\parallel n_i n_j+D_\perp(\delta_{ij}-n_in_j)$, with $\hat{n}=(\cos{\theta},\sin{\theta})$, the unit vector pointing along the particle’s major axis, $v$ is the propulsion velocity of the anisotropic particle along its major axis. Taking a Fourier transform,

\begin{equation}
    \hat{P}_0(k,\theta,t|\theta_0)\equiv\int dx e^{-i\textbf{k}.\textbf{x}} P_0(\textbf{x},\theta,t|\theta_0)
\end{equation}

The transformed equation becomes,

\begin{equation}
    \partial_t\hat{P}_0(k,\theta,t|\theta_0)=D_\theta \partial_\theta^2\hat{P}_0(k,\theta,t|\theta_0)+k[\mathbf{D}.k]\hat{P}_0(k,\theta,t|\theta_0)-ivk\cos{\theta}\hat{P}_0(k,\theta,t|\theta_0)
\end{equation}

To simplify the analysis, we focus on the steady-state behavior in the $x$-direction. Accordingly, we consider a wave vector $\textbf{k}=ke_x$ with $e_x$ the unit vector in the $x$-direction. Under these conditions, the marginal distribution in the xx-direction satisfies:

\begin{equation}
    \partial_t\hat{P}_0(k,\theta,t|\theta_0)=\Bigg[D_\theta \partial_\theta^2-k^2\Big(\Bar{D}+\frac{\Delta D}{2}\cos{2\theta}\Big)\Bigg]\hat{P}_0(k,\theta,t|\theta_0)-ivk\cos{\theta}\hat{P}_0(k,\theta,t|\theta_0)
\end{equation}

A straightforward interpretation of this equation is that the diffusivity experienced by the particle in the $x$-direction, given by $\Bar{D}+\frac{\Delta D}{2}\cos{2\theta}$ , varies depending on the particle's orientation. To derive an analytical solution that sheds light on the steady-state shape under resetting, we adopt a perturbative approach. Specifically, we expand the solution in powers of the dimensionless asymmetry parameter $\epsilon\equiv\frac{\Delta D}{\Bar{D}}$. The marginal Smoluchowski-Perrin equation can then be expressed as:

\begin{equation}
    \partial_t\hat{P}_0(k,\theta,t|\theta_0)=D_\theta\partial_\theta^2\hat{P}_0(k,\theta,t|\theta_0)-k^2\Bar{D}\hat{P}_0(k,\theta,t|\theta_0)-\epsilon k^2\Bar{D}\frac{\cos{2\theta}}{2}\hat{P}_0(k,\theta,t|\theta_0)-ivk\cos{\theta}\hat{P}_0(k,\theta,t|\theta_0)
\end{equation}

We expand the solution in powers of the asymmetry as

\begin{equation}
    \hat{P}_0(k,\theta,t|\theta_0)=\sum_{n=0}^{\infty}p_n(k,\theta,t|\theta_0)\epsilon^n
\end{equation}

The function $p_n(k,\theta,t|\theta_0)$ represents the coefficient of the $n$-th term in the power series expansion. Substituting this ansatz into the Smoluchowski-Perrin equation results in a set of coupled equations that govern these coefficients. These equations are derived by equating terms with the same powers of the asymmetry parameter $\epsilon$, leading to a hierarchy of equations that must be solved sequentially to determine the behavior of the system under the perturbative expansion.

\begin{equation}
    \partial_tp_0(k,\theta,t|\theta_0)=D_\theta \partial_\theta^2p_0(k,\theta,t|\theta_0)-k^2\bar{D}p_0(k,\theta,t|\theta_0)-ivk\cos{\theta}p_0(k,\theta,t|\theta_0)
\end{equation}
\begin{equation}
     \partial_tp_n(k,\theta,t|\theta_0)=D_\theta \partial_\theta^2p_n(k,\theta,t|\theta_0)-k^2\bar{D}p_n(k,\theta,t|\theta_0)-\frac{k^2\bar{D}\cos{2\theta}}{2}p_{n-1}(k,\theta,t|\theta_0)-ivk\cos{\theta}p_n(k,\theta,t|\theta_0)
    \end{equation}

First, it is important to observe that $p_0(k,\theta,t)$ corresponds to the solution characteristic of a symmetric colloid. Since the focus lies on determining the spatial steady-state distribution to first order in the perturbative expansion, we integrate over the angular degree of freedom for $n=0$ and $n=1$. This integration results in expressions that describe the marginal steady-state distributions, incorporating the leading-order effects of anisotropy in the system. The integration effectively reduces the problem to a spatial one while capturing the perturbative corrections introduced by the asymmetry parameter.

\begin{equation}
    \partial_tp_0(k,t|\theta_0)=-k^2\Bar{D}p_0(k,t|\theta_0)-ivk\cos{\theta_0}e^{-D_\theta t}p_0(k,t|\theta_0)
\end{equation}
\begin{equation}
    \partial_tp_1(k,t|\theta_0)=-k^2\Bar{D}p_1(k,t|\theta_0)-\frac{k^2\bar{D}}{2}\cos{2\theta_0}e^{-4D_\theta t}p_0(k,t|\theta_0)-ivk\cos{\theta_0}e^{-D_\theta t}p_1(k,t|\theta_0)
\end{equation}

Here the zeroth order solution can be written as

\begin{equation}
    p_0(k,t)=e^{-k^2\Bar{D}t}e^{-\frac{ivk\cos{\theta_0}}{D_\theta}}e^{\frac{ivk\cos{\theta_0}}{D_\theta}e^{-D_\theta t}}
\end{equation}

If we take an approximation that the propulsion velocity $v$ is very small and in long time $e^{-D_\theta t}$ term also goes to very smaller, so we can expand the exponential term upto first order and we get,

\begin{equation}
    p_0(k,t)\approx e^{-k^2\Bar{D}t}e^{-\frac{ivk\cos{\theta_0}}{D_\theta}}\Bigg(1+\frac{ivk\cos{\theta_0}}{D_\theta}e^{-D_\theta t}\Bigg)
\end{equation}

Performing a Laplace transform of the above equation followed by a Fourier inverse transform, we obtain

\begin{equation}
    \Tilde{p}_0(x,s|\theta_0)=\frac{\beta(s)}{2s}e^{-\beta(s)|(x-\frac{v\cos{\theta_0}}{D_\theta})|}-\frac{v\cos{\theta_0}}{2D_\theta\Bar{D}}e^{-\beta(s+D_\theta)|(x-\frac{v\cos{\theta_0}}{D_\theta})|}
    \label{first}
\end{equation}

here $\beta(s)=\sqrt{s/\Bar{D}}$ is the inverse lengthscale.

The integration in the equation for the first-order correction can be performed exactly, yielding an explicit expression for the first-order term in the perturbative expansion.

\begin{equation}
    \partial_tp_1(k,t|\theta_0)=-k^2\bar{D}p_1(k,t|\theta_0)-\frac{k^2\Bar{D}}{2}\cos{2\theta_0}e^{-(4D_\theta+k^2\Bar{D}) t}e^{-\frac{ivk\cos{\theta_0}}{D_\theta}}\Bigg(1+\frac{ivk\cos{\theta_0}}{D_\theta}e^{-D_\theta t}\Bigg)-ivk\cos{\theta_0}e^{-D_\theta t}p_1(k,t|\theta_0)
\end{equation}

Performing a Laplace transform results in

\begin{equation}
 \Tilde{\hat{p}}_1(k,s)=-\frac{k^2\bar{D}\cos{2\theta_0}e^{-\frac{ivk\cos{\theta_0}}{D_\theta}}}{2D_\theta}\frac{D_\theta\big(s+k^2\Bar{D}+5D_\theta\big)+ivk\cos{\theta_0}\big(s+k^2\Bar{D}+4D_\theta)}{\big(s+k^2\Bar{D}+4D_\theta\big)\big(s+k^2\Bar{D}+5D_\theta\big)\bigg(s+k^2\Bar{D}+\big(s+D_\theta\big)ivk\cos{\theta_0}\bigg)}   
\end{equation}

By inverting the Fourier transform, we ultimately derive the first-order correction to the propagator as

\begin{equation}
    \begin{split}
        &\Tilde{p}_1(x,s|\theta_0)=\frac{\cos{2\theta_0}}{4}\Bigg[\frac{e^{\sqrt{\frac{s+4D_\theta}{\Bar{D}}}(x-\frac{v\cos{\theta_0}}{D_\theta})}}{4D_\theta\sqrt{\frac{\Bar{D}}{s+4D_\theta}}-v\cos{\theta_0}(s+D_\theta)}-\frac{e^{-\sqrt{\frac{s+4D_\theta}{\Bar{D}}}(x-\frac{v\cos{\theta_0}}{D_\theta})}}{4D_\theta\sqrt{\frac{\Bar{D}}{s+4D_\theta}}+v\cos{\theta_0}(s+D_\theta)}\Bigg]+\frac{v\cos{\theta_0}\cos{2\theta_0}(s+5D_\theta)}{4D_\theta}\\
        &\Bigg[\frac{e^{\sqrt{\frac{s+5D_\theta}{\Bar{D}}}(x-\frac{v\cos{\theta_0}}{D_\theta})}}{5D_\theta\Bar{D}-v\cos{\theta_0}(s+D_\theta)\sqrt{\Bar{D}(s+D_\theta)}}+\frac{e^{-\sqrt{\frac{s+5D_\theta}{\Bar{D}}}(x-\frac{v\cos{\theta_0}}{D_\theta})}}{5D_\theta\Bar{D}+v\cos{\theta_0}(s+D_\theta)\sqrt{\Bar{D}(s+D_\theta)}}\Bigg]\\
        &+\frac{\cos{2\theta_0}}{2}\Bigg[\frac{\Big(\alpha^2(s,v,D_\theta)+2s\Bar{D}-\alpha(s,v,D_\theta)\sqrt{\alpha^2(s,v,D_\theta)+4s\Bar{D}}\Big)e^{-\frac{1}{2\Bar{D}}\Big(-   \alpha(s,v,D_\theta)+\sqrt{\alpha^2(s,v,D_\theta)+4s\Bar{D}}\Big)(x-\frac{v\cos{\theta_0}}{D_\theta})}}{\sqrt{\alpha^2(s,v,D_\theta)+4s\Bar{D}}\Big[8D_\theta\Bar{D}-\alpha(s,v,D_\theta)(\alpha(s,v,D_\theta)-\sqrt{\alpha^2(s,v,D_\theta)}+4s\Bar{D})\Big]}\\
        &-\frac{\Big(\alpha^2(s,v,D_\theta)+2s\Bar{D}+\alpha(s,v,D_\theta)\sqrt{\alpha^2(s,v,D_\theta)+4s\Bar{D}}\Big)e^{\frac{1}{2\Bar{D}}\Big(\alpha(s,v,D_\theta)+\sqrt{\alpha^2(s,v,D_\theta)+4s\Bar{D}}\Big)(x-\frac{v\cos{\theta_0}}{D_\theta})}}{\sqrt{\alpha^2(s,v,D_\theta)+4s\Bar{D}}\Big[8D_\theta\Bar{D}-\alpha(s,v,D_\theta)(\alpha(s,v,D_\theta)+\sqrt{\alpha^2(s,v,D_\theta)}+4s\Bar{D})\Big]}\Bigg]\\
        &+\frac{v\cos{\theta_0}\cos{2\theta_0}}{8D_\theta\Bar{D}}\Bigg[\frac{\Big(4\alpha^3+3\alpha^2\sqrt{\alpha^2+4s\Bar{D}}+12\alpha s\Bar{D}+(\alpha^2+4s\Bar{D})^{3/2}\Big)e^{\frac{1}{2\Bar{D}}\Big(\alpha(s,v,D_\theta)+\sqrt{\alpha^2(s,v,D_\theta)+4s\Bar{D}}\Big)(x-\frac{v\cos{\theta_0}}{D_\theta})}}{\sqrt{\alpha^2+4s\Bar{D}}\Big(10D_\theta\Bar{D}-\alpha(\alpha+\sqrt{\alpha^2+4s\Bar{D}})\Big)}\\
        &-\frac{\Big(4\alpha^3-3\alpha^2\sqrt{\alpha^2+4s\Bar{D}}+12\alpha s\Bar{D}-(\alpha^2+4s\Bar{D})^{3/2}\Big)e^{-\frac{1}{2\Bar{D}}\Big(-\alpha(s,v,D_\theta)+\sqrt{\alpha^2(s,v,D_\theta)+4s\Bar{D}}\Big)(x-\frac{v\cos{\theta_0}}{D_\theta})}}{\sqrt{\alpha^2+4s\Bar{D}}\Big(10D_\theta\Bar{D}-\alpha(\alpha-\sqrt{\alpha^2+4s\Bar{D}})\Big)}\Bigg]
    \end{split}
\end{equation}
here $\alpha(s,v,D_\theta)=v\cos{\theta_0}(s+D_\theta)$

The spatial probability distribution up to the first order correction without resetting is the combination of both $\Tilde{p}_0(x,s)$ and $\Tilde{p}_1(x,s)$ 

\begin{equation}
    \Tilde{P}_0(x,s|\theta_0)=\Tilde{p}_0(x,s|\theta_0)+\epsilon\Tilde{p}_1(x,s|\theta_0)
    \label{100}
\end{equation}
The term $ \Tilde{P}_0(x,s|\theta_0)$ represents a normalized probability distribution. We can now readily obtain the steady state distribution under the effect of resetting by using the renewal equation Eq.(\ref{12})

\begin{equation}
    P_r(x|0,\theta_0)=r\int_{0}^{\infty}d\tau e^{-r\tau} P_0(x,\tau|0,\theta_0)=r\Tilde{P}_0(x,r|0,\theta_0)
    \label{101}
\end{equation}
Substituting the value of $\Tilde{P}_0(x,s)$ from Eq.(\ref{100}) in the Eq.(\ref{101}) and replacing $s$ as $r$, we get

\begin{equation}
    \begin{split}
        &\Tilde{P}_r(x|\theta_0)=\frac{\beta(r)}{2}e^{-\beta(r)|(x-\frac{v\cos{\theta_0}}{D_\theta})|}-\frac{v\cos{\theta_0}}{2D_\theta\Bar{D}}e^{-\beta(r+D_\theta)|(x-\frac{v\cos{\theta_0}}{D_\theta})|}+\frac{\epsilon\cos{2\theta_0}}{4}\Bigg[\frac{re^{\beta(r+4D_\theta)(x-\frac{v\cos{\theta_0}}{D_\theta})}}{4D_\theta\sqrt{\frac{\Bar{D}}{r+4D_\theta}}-v\cos{\theta_0}(r+D_\theta)}\\
        &-\frac{re^{-\beta(r+4D_\theta)(x-\frac{v\cos{\theta_0}}{D_\theta})}}{4D_\theta\sqrt{\frac{\Bar{D}}{r+4D_\theta}}+v_0\cos{\theta_0}(r+D_\theta)}\Bigg]+\frac{\epsilon v\cos{\theta_0}\cos{2\theta_0}(r+5D_\theta)}{4D_\theta}\Bigg[\frac{re^{\beta(r+5D_\theta)(x-\frac{v\cos{\theta_0}}{D_\theta})}}{5D_\theta\Bar{D}-v\cos{\theta_0}(r+D_\theta)\sqrt{\Bar{D}(r+D_\theta)}}\\
        &+\frac{re^{-\beta(r+5D_\theta)(x-\frac{v\cos{\theta_0}}{D_\theta})}}{5D_\theta\Bar{D}+v\cos{\theta_0}(r+D_\theta)\sqrt{\Bar{D}(r+D_\theta)}}\Bigg]\\
        &+\frac{\epsilon r\cos{2\theta_0}}{2}\Bigg[\frac{\Big(\alpha^2(r,v,D_\theta)+2r\Bar{D}-\alpha(r,v,D_\theta)\sqrt{\alpha^2(r,v,D_\theta)+4r\Bar{D}}\Big)e^{-\frac{1}{2\Bar{D}}\Big(-   \alpha(r,v,D_\theta)+\sqrt{\alpha^2(r,v,D_\theta)+4r\Bar{D}}\Big)(x-\frac{v\cos{\theta_0}}{D_\theta})}}{\sqrt{\alpha^2(r,v,D_\theta)+4r\Bar{D}}\Big[8D_\theta\Bar{D}-\alpha(r,v,D_\theta)(\alpha(r,v,D_\theta)-\sqrt{\alpha^2(r,v,D_\theta)}+4r\Bar{D})\Big]}\\
        &-\frac{\Big(\alpha^2(r,v,D_\theta)+2r\Bar{D}+\alpha(r,v,D_\theta)\sqrt{\alpha^2(r,v,D_\theta)+4r\Bar{D}}\Big)e^{\frac{1}{2\Bar{D}}\Big(\alpha(r,v,D_\theta)+\sqrt{\alpha^2(r,v,D_\theta)+4r\Bar{D}}\Big)(x-\frac{v\cos{\theta_0}}{D_\theta})}}{\sqrt{\alpha^2(r,v,D_\theta)+4r\Bar{D}}\Big[8D_\theta\Bar{D}-\alpha(r,v,D_\theta)(\alpha(r,v,D_\theta)+\sqrt{\alpha^2(r,v,D_\theta)}+4r\Bar{D})\Big]}\Bigg]
    \end{split}
    \label{212}
\end{equation}

There are several noteworthy aspects to consider in the case of an anisotropic active Brownian particle (ABP). The steady-state probability distribution at zeroth order is defined by a characteristic length scale, represented by $\beta(r)$, and the particle's activity introduces an additional length scale, $\beta(r + D_\theta)$, which is influenced by the resetting rate. The distribution exhibits a peak at $x = 0$ as a result of the restrictive influence exerted by resetting at the origin. This peak value rises with an increase in the resetting rate $r$, as anticipated. 

When the anisotropy of the particle is considered, the solution becomes dependent on the changes in length scales and is influenced by the propulsion velocity, reset rate, rotational diffusivity and the initial orientation angle $\theta_0$, introduced by the anisotropy of the particle. Interestingly, the rotational dynamics influence the spatial steady-state distribution through the coupling between translational and rotational degrees of freedom, an effect that endures even in the long-time regime.

\section{EXACT COMPUTATION OF THE
MOMENTS FOR THE POSITION RESETTING}\label{appB}

The second moment of $x$ for the anisotropic ABP putting $n=2$ in Eq.(\ref{1})
\begin{equation}
\begin{split}
\langle x^2(t)|\theta_0\rangle_r&=e^{-rt}\Big[2\Bar{D}t+\frac{\Delta D\cos{2\theta_0}}{4D_\theta}(1-e^{-4D_\theta t})+\frac{v_0^2\cos{2\theta_0}}{12D_\theta^2}(3-4e^{-D_\theta t}+e^{-4D_\theta t})+\frac{v_0^2}{D_\theta^2}(D_\theta t+e^{-D_\theta t}-1)\Big]\\
&+r\int_{0}^{t}d\tau e^{-r\tau}\int d\theta^\prime\frac{e^{-\frac{(\theta^\prime-\theta_0)^2}{4D_\theta(t-\tau)}}}{\sqrt{4\pi D_\theta(t-\tau)}}\Big[2\Bar{D}\tau+\frac{\Delta D\cos{2\theta^\prime}}{4D_\theta}(1-e^{-4D_\theta\tau})+\frac{v_0^2\cos{2\theta_0}}{12D_\theta^2}(3-4e^{-D_\theta\tau}+e^{-4D_\theta\tau})\\
&+\frac{v_0^2}{D_\theta^2}(D_\theta\tau+e^{-D_\theta\tau}-1)\Big]
\end{split}
\end{equation}
We will use the following properties to calculate the above integral
\begin{equation}
\begin{split}
&\langle\cos{2\theta(t)}|\theta_0\rangle=Re\langle e^{i2\theta}|\theta_0\rangle=Re\Bigg[\int d\theta e^{i2\theta}\frac{e^{-\frac{(\theta-\theta_0)^2}{4D_\theta(t-\tau)}}}{\sqrt{4\pi D_\theta(t-\tau)}}\Bigg]=\cos{2\theta_0}e^{-4D_\theta(t-\tau)}\\
&\int d\theta\frac{e^{-\frac{(\theta-\theta_0)^2}{4D_\theta(t-\tau)}}}{\sqrt{4\pi D_\theta(t-\tau)}}=1
\end{split}
\end{equation}

\section{EXACT COMPUTATION OF THE
MOMENTS FOR THE ORIENTATION RESETTING}\label{appC}
The Langevin Eqs.(\ref{lab}) can be formally integrated to write
\begin{equation}
    \begin{split}
        &x(t)=\int_{0}^{t}dt^\prime\eta_x(t^\prime)+\int_{0}^{t}dt^\prime v_0\cos{\theta}(t^\prime)\\
        &y(t)=\int_{0}^{t}dt^\prime\eta_y(t^\prime)+\int_{0}^{t}dt^\prime v_0\sin{\theta}(t^\prime)
    \end{split}
    \label{8}
\end{equation}
Taking the statistical average over all possible trajectories of $\theta$, we obtain the following expression 
\begin{equation}
    \begin{split}
       &\langle x(t)\rangle= \int_{0}^{t}dt^\prime v_0\Big\langle\cos{\theta}(t^\prime)\Big\rangle\\
       &\langle y(t)\rangle= \int_{0}^{t}dt^\prime v_0\Big\langle\sin{\theta}(t^\prime)\Big\rangle
    \end{split}
    \label{5}
\end{equation}
Let $\mathcal{P}(\theta, t|\theta^\prime, t^\prime )$ represent the probability that the orientation takes the value $\theta$ at time $t$ given that it was $\theta^\prime$ at an earlier
time $t^\prime$. The evolution of this probability distribution satisfies a renewal equation, which captures the dynamics of the system in the presence of resetting events.

\begin{equation}
    \mathcal{P}(\theta, t|\theta^\prime, t^\prime )=e^{-r(t-t^\prime)}\mathcal{P}_0(\theta, t|\theta^\prime, t^\prime )+r\int_{0}^{(t-t^\prime)}d\tau e^{-r\tau}\mathcal{P}_0(\theta,\tau|0,0)
    \label{6}
\end{equation}
where $\mathcal{P}_0(\theta, t|\theta^\prime, t^\prime )$ denotes the propagator for the standard
Brownian motion, given as
\begin{equation}
    \mathcal{P}_0(\theta, t|\theta^\prime, t^\prime )=\frac{1}{\sqrt{4\pi D_\theta(t-t^\prime)}}exp\Bigg[-\frac{(\theta-\theta^\prime)^2}{4D_\theta(t-t^\prime)}\Bigg]
\end{equation}

The averages appearing on the right-hand side of the Eq.(\ref{5}) can be
computed using the renewal equation Eq.(\ref{6}) for $ \mathcal{P}(\theta, t)$. We get,
\begin{equation}
    \langle\cos{\theta}(\tau)\rangle=\int_{-\infty}^{+\infty}d\theta\cos{\theta}\mathcal{P}(\theta,\tau)=\frac{D_\theta}{r+D_\theta}e^{-(r+D_\theta)\tau}+\frac{r}{r+D_\theta}
\end{equation}
and $\langle\sin{\theta}(s)\rangle=0$. Using the above expression in Eq.(\ref{5}) we can calculate the exact expression for the mean of $x$ position as,
\begin{equation}
   \langle x(t)|\theta_0\rangle=\frac{v_0}{r+D_\theta}\Bigg[rt+\frac{D_\theta}{r+D_\theta}\Big(1-e^{-(r+D_\theta)t}\Big)\Bigg]
    \label{7}
\end{equation}
And $\langle y(t)\rangle=0$
\subsection{Variance of $x(t )$ and $y(t )$}
We get from Eq.(\ref{8}) 

\begin{equation}
    \begin{split}
        &\langle x^2(t)|\theta_0\rangle_{or}=v_0^2\int_{0}^{t}d\tau\int_{0}^{t}d\tau^\prime\Big\langle\cos{\theta}(\tau)\cos{\theta}(\tau^\prime)\Big\rangle+\int_{0}^{t}d\tau\int_{0}^{t}d\tau^\prime\Big\langle\eta(\tau)\eta(\tau^\prime)\Big\rangle\\
        &\langle y^2(t)|\theta_0\rangle_{or}=v_0^2\int_{0}^{t}d\tau\int_{0}^{t}d\tau^\prime\Big\langle\sin{\theta}(\tau)\sin{\theta}(\tau^\prime)\Big\rangle+\int_{0}^{t}d\tau\int_{0}^{t}d\tau^\prime\Big\langle\eta(\tau)\eta(\tau^\prime)\Big\rangle
    \end{split}
    \label{B7}
\end{equation}

To compute the position moments, it is essential to first evaluate the two-time auto-correlations in the governing equations. Beginning with the two-time correlation of $\cos{\theta}$; for $\tau>\tau^\prime$
we get,
\begin{equation}
\begin{split}
    C(\tau,\tau^\prime)&\equiv\Big\langle\cos{\theta}(\tau)\cos{\theta}(\tau^\prime)\Big\rangle\\
    &=\int d\theta d\theta^\prime\cos{\theta}\cos{\theta^\prime}\mathcal{P}(\theta,\tau|\theta^\prime,\tau^\prime)\mathcal{P}(\theta,\tau^\prime|0,0)
    \end{split}
\end{equation}
where the propagator $\mathcal{P}(\theta,\tau|\theta^\prime,\tau^\prime)$ satisfies the renewal Eq.(\ref{6}). Using Eq.(\ref{6}) in the above equation and performing the integrals, we get, for $\tau>\tau^\prime$,
\begin{equation}
    \begin{split}
        C(\tau,\tau^\prime)&=\frac{r^2}{(r+D_\theta)^2}+\frac{2D_\theta}{r+4D_\theta}e^{-(r+D_\theta)\tau-3D_\theta \tau^\prime}+\frac{rD_\theta}{(r+D_\theta)^2}\Big(e^{-(r+D_\theta)\tau^\prime}-e^{-(r+D_\theta)\tau}\Big)\\
        &+\frac{D_\theta^2(5r+2D_\theta)}{(r+4D_\theta)(r+D_\theta)^2}e^{-(r+D_\theta)(\tau-\tau^\prime)}
    \end{split}
    \label{B9}
\end{equation}
To calculate the second term containing $\langle\eta(\tau)\eta(\tau^\prime)\rangle$ in RHS of Eq.(\ref{B7}), we approach as 
\begin{equation}
    \begin{split}
        \int_{0}^{t}d\tau\int_{0}^{t}d\tau^\prime\langle\eta(\tau)\eta(\tau^\prime)\rangle=2k_BT\int_{0}^{t}d\tau\int_{0}^{t}d\tau^\prime\Bigg[\bar{\Gamma}+\frac{\Delta\Gamma}{2}\langle\cos{2\theta}(\tau^\prime)|\theta_0\rangle_{or}\Bigg]
    \end{split}
    \label{B10}
\end{equation}
To determine $\langle\cos{\theta}(\tau^\prime)|\theta_0\rangle_{or}$ under resetting conditions, we apply the renewal equation specifically for $\theta$, which simplifies to the following form:

\begin{equation}
    P_r(\theta,t|\theta_0)=e^{-rt}P_0(\theta,t|\theta_0)+r\int_{0}^{t}d\tau e^{-r\tau}P_0(\theta,\tau|\theta_0)
\end{equation}
By multiplying the governing equation with $\cos{2\theta}$ and integrating, we obtain
\begin{equation}
    \begin{split}
        \langle\cos{2\theta}(\tau)|\theta_0\rangle_{or}&=e^{-rt}\langle\cos{2\theta}(\tau)|\theta_0\rangle_0+r\int_{0}^{t}dt^\prime e^{-rt^\prime}\langle\cos{2\theta}(t^\prime)|\theta_0\rangle_0\\
        &=\cos{2\theta_0}e^{-(r+4D_\theta)t}+\frac{r\cos{2\theta_0}}{r+4D_\theta}\Big(1-e^{-(r+4D_\theta)t}\Big)
    \end{split}
\end{equation}
Using the above expression in Eq.(\ref{B10}) we find,
\begin{equation}
    \begin{split}
        \int_{0}^{t}d\tau\int_{0}^{t}d\tau^\prime\langle\eta(\tau)\eta(\tau^\prime)\rangle&=2k_BT\int_{0}^{t}d\tau^\prime\Big[\bar{\Gamma}+\frac{\Delta\Gamma}{2}\langle\cos{2\theta}(\tau^\prime)|\theta_0\rangle_{or}\Big]\\
        &=2k_BT\Bigg[\Big(\bar{\Gamma}+\frac{r\Delta\Gamma\cos{2\theta_0}}{2(r+4D_\theta)}\Big)t-\frac{2D_\theta\Delta\Gamma\cos{2\theta_0}}{(r+4D_\theta)^2}(1-e^{-(r+4D_\theta)t})\Bigg]
    \end{split}
    \label{B13}
\end{equation}

Repeating the same exercise for $\sin{\theta}$, we get
\begin{equation}
    \Big\langle\sin{\theta}(\tau)\sin{\theta}(\tau^\prime)\Big\rangle=\frac{2D_\theta}{r+4D_\theta}e^{-(r+D_\theta)(\tau-\tau^\prime)}\Big[1-e^{-(r+4D_\theta)\tau^\prime}\Big]
\end{equation}

Similarly, the second moment of the $y$-component is calculated using the renewal framework and incorporating the effects of resetting and anisotropy

\begin{equation}
    \begin{split}
        \langle y^2(t)|\theta_0\rangle_{or}&=\Bigg[2\Bigg(\bar{D}-\frac{r\Delta D}{r+4D_\theta}\Bigg)t-\frac{4D_\theta\Delta D}{(r+4D_\theta)^2}\Big(1-e^{-(r+4D_\theta)t}\Big)\Bigg]+\frac{4v_0^2D_\theta t}{(r+4D_\theta)(r+D_\theta)}-\frac{4v_0^2D_\theta(2r+5D_\theta)}{(r+4D_\theta)^2(r+D_\theta)^2}\\
        &+\frac{4v_0^2}{3}\Bigg(\frac{e^{-(r+D_\theta)t}}{(r+D_\theta)^2}-\frac{e^{-(r+4D_\theta)t}}{(r+4D_\theta)^2}\Bigg)
    \end{split}
    \label{C15}
\end{equation}

\end{appendix}
\end{document}